\documentclass[traditabstract]{aa} 
\usepackage{txfonts}
\usepackage{longtable}
\usepackage{graphicx}
\usepackage{natbib}
\usepackage{colortbl}
\def\micron{\hbox{$\mu$m}}
\begin{document}

\title{The VISIR@VLT Mid-IR view of 47Tuc
\thanks{Based on data obtained at the ESO/UT3 proposal 084.D-0721(A).%
}}

\subtitle{A further step in solving the puzzle of RGB mass loss}

\author{Y.    Momany\inst{1,2}    \and    I.   Saviane\inst{1}    \and
  A. Smette\inst{1} \and A.  Bayo\inst{1} \and L. Girardi\inst{2} \and
  G. Marconi\inst{1} \and A. P. Milone\inst{3,4} \and A. Bressan\inst{2,5}}

\institute{European  Southern  Observatory,  Alonso de  Cordova  3107,
  Santiago,                                                      Chile.
  \email{ymomany,isaviane,gmarconi,asmette,abayo@eso.org}  \and  INAF,
  Oss.   Astronomico  di  Padova, Vicolo  dell'Osservatorio  5,I-35122
  Padova,                                                        Italy.
  \email{yazan.almomany,leo.girardi,alessandro.bressan@oapd.inaf.it}
  \and  Instituto de  Astrofísica  de Canarias,  La Laguna,  Tenerife,
  Spain   \email{milone@iac.es}   \and   Department  of   Astrophysic,
  University of La Laguna, E-38200 La Laguna, Tenerife, Canary Island,
  Spain \and SISSA, via Bonomea, 265, 34136 Trieste, Italy.}

\date{Received May 10, 2011; accepted October 10, 2011}

\abstract
{There  is an ongoing debate regarding  the onset luminosity
  of dusty mass loss in population-II red giant stars.
  We   present  VISIR@VLT  mid-infrared   (MIR)  $\mathrm{8.6\micron}$
  imaging of 47Tuc, the centre of attention of a number of space-based
  Spitzer observations and studies.
  The VISIR  high-resolution (diffraction  limited) observations allow
  excellent  matching  to  existing  optical  Hubble  space  telescope
  catalogues. The  optical-MIR coverage of the  inner $1\farcm15$ core
  of  the  cluster   provide  the  cleanest  possible  blending-free
  sampling of the upper 3 magnitudes of the giant branch.
  Our diagrams  show no  evidence of faint  giants with  MIR-excess. A
  combined  near/mid-infrared diagram  additionally confirms  the near
  absence of dusty red giants.
%
  Dusty  red giants  and asymptotic  giant stars  are confined  to the
  47Tuc  long-period variable  population.  In  particular,  dusty red
  giants  are  limited  to  the  upper  one  $N_{\mathrm{8.6\micron}}$
  magnitude below  the giant  branch tip.  This  particular luminosity
  level  corresponds to  $\sim1000L_{\odot}$, which  was  suggested in
  previous  determinations  to mark  the  onset  of  dusty mass  loss.
  Interestingly,  starting from  this luminosity  level we  detected a
  gradual  deviation  between  the  colours  of  red  giants  and  the
  theoretical isochrones.}
\keywords{infrared: stars -- stars: individual (Population II) -- stars: winds,
outflows -- stars: mass-loss -- globular clusters: individual: 47 Tucanae (NGC104, 47Tuc) }

\maketitle

\section{Introduction}

During the final stages of their red giant and asymptotic giant branch
(hereafter RGB,  AGB respectively) ascent, stars  suffer a significant
mass loss.  Direct evidence of  this process in AGBs has been recently
provided by spectroscopic mid-infrared (MIR) observations with ISO and
SPITZER.
The common view  is that a typical turnoff star  in a globular cluster
of  $\sim0.8M_{\odot}$  may  loose  up  to $\sim  30\%$  of  its  mass
(\citealt{caloi08}, and  references therein), i.e.  $\sim0.2M_{\odot}$
during the RGB and $\sim0.1M_{\odot}$ during the AGB phases.
%
Mass loss  along the  RGB is an  important phenomenon because  it will
eventually  alter  (i) the  horizontal  branch  morphology; (ii)  the
 RR Lyrae pulsational  properties;  (iii) the  ratios of  O-rich to
C-rich AGB  star production; (iv)  the post-AGB and  planetary nebula
chemistry,    and   lastly    (v)   the    mass   of    white   dwarfs
(\citealt{mcdonald09}).
However, a consistent theoretical  picture of the mass loss phenomenon
is still lacking and ultimately the inability to understand/model this
important   process,   which   is    the   major   polluter   of   the
interstellar-medium (ISM), hinders  our interpretation (e.g.  inferred
stellar masses,  ages and  metallicities) of high-redshift unresolved
galaxies.
In the context of stellar systems such as globular and open clusters, the
interest in  the RGB/AGB  mass loss mechanism  has gained a  whole new
dimension of  importance in  the past six  years. 
 Indeed,  the discovery of  multiple and discrete  main sequences
  {\em  with different  helium  contents} in  many globular  clusters
  (\citealt{piotto05};  \citealt{renzini08}) is nowadays  explained by
  invoking    mass loss  from  the  cluster's  primary generation  of
intermediate-mass  red   giants  and   AGB  stars that  enriched  the
interstellar medium, which later  ``gave birth'' to subsequent generations
of chemically enriched stars  (see \citealt{eug09}, and references
  therein  for  an  analysis  of multiple  populations  and  chemical
  inhomogeneities in 15 Galactic globular clusters).

Early qualitative  estimates of mass loss  were provided theoretically
by  \citet{rood73},  who  noticed  that  without  mass  loss  the  blue
horizontal branch (HB) stars of  even metal-poor clusters could not be
reproduced by the models.
At  the same time  \citet{reimers75a} provided  observational evidence
from Population~I  giants, and parameterized mass loss  with the widely
used so-called Reimers's formula.   The formula is still used today,
although  its  original  theoretical  motivation  needs to be  somewhat
relaxed to  accommodate the presence  of multiple populations  that may
reach  a very high  helium content  which then leads to  hot  extended horizontal
branches (\citealt{fagotto94}; \citealt{dantona08}).

An important  outcome of the Reimers'  formula is that  mass loss only
becomes  significant near  the tip  of the  RGB.  This  view  has been
recently challenged by \citet{origlia07},  who analysed red sources in
47Tuc and concluded that the  dependence of the derived mass loss rate
on  luminosity   is  considerably   flatter  than  predicted   by  the
\citet{reimers75a,reimers75b}  formulation.   This  leads to  several
important  consequences  for  the  subsequent  evolution  of  low-  and
intermediate-mass stars.

\citet{origlia07}  based  their  conclusions  on  the  analysis  of  a
combined  near-  and mid-infrared  (NIR+MIR) colour-magnitude  diagram
(CMD) of  47Tuc based on IRAC2@ESO2.2m and  SOFI@ESO-NTT photometry in
the $J$- and $K$-bands and Spitzer/IRAC photometry in the MIR.
In   their   $M_{{\rm   bol}}$   {\it   vs.}    $(K-8\micron)_{\circ}$
colour-magnitude diagrams the authors detected a  MIR excess for over 100 red
giants. This excess  was interpreted as the results  of dust formation
around  these stars.   These  stars  form a  sequence  that is  almost
parallel to the ``normal'' RGB  (c.f.  their Fig.  2) and reaches the
horizontal branch level.

The   results   of    \citet{origlia07}   were   soon   challenged   by
\citet{boyer08}.  According to the latter's Spitzer $\omega$Cen study,
significant dusty  mass loss can  occur only at  or near the  RGB tip
 (as originally  found in \citealt{origlia02}). This is   at odds
with the findings  of \citet{origlia07} for 47Tuc, where  mass loss was
detected down to $\sim4~K$-magnitudes below the RGB tip (TRGB).  These
discrepancies could  not be attributed to  differences in metallicity
between the two clusters.
Indeed,  \citet{boyer08}  estimated that  at  least  25\%  of the  red
sources  in  $\omega$Cen  should  have  red colours  because  of  {\em
  blending  effects} at  $24$\micron.  Therefore,  they suggested  that the
higher   crowding  conditions   of   the  47Tuc   core  (core   radius
$r_{\rm c}=0\farcm4$~vs.   the  $1\farcm4$   of  $\omega$Cen)  might  have
affected an even higher percentage of red stars.  In particular, Boyer
et al. (2008)  suggested that the absence of red  sources in the outer
and  less  crowded  47Tuc  region  \citep[see][Fig.~2]{origlia07}  may
indicate that more crowded  core regions could  induce the
MIR-excess.

Subsequently, \citet{boyer10} supported  their claim by studying 47Tuc
directly.   Lacking NIR  data,  they  showed CMDs  based  on the  same
independently  reduced Spitzer  data  of Origlia  et  al.,  which  they
complemented  with  photometry from  the  SAGE-SMC  survey.  In  their
$M_{{\rm  {\rm  bol}}}$  {\it vs.}   $(3.6\micron  -8\micron)_{\circ}$
diagrams,  no colour-excess  stars  were observed for  objects more  than
$1$~mag fainter than the TRGB.
Therefore they concluded that at $\sim1$ mag.  below the RGB tip there
were  {\em  no} red  giant  stars  producing  dust. Once  again,  they
attributed the Origlia et al.  results to stellar blending and imaging
artefacts, and  consequently, cast doubt on the  mass loss law derived
by Origlia et al.

\citet{origlia10}  responded to  the  Boyer et  al.
skepticism.  Their  principal argument  was summarized in  their Fig.~1,
where  they showed  that an  excess  was  much more  obvious in  the
$K-8\micron$ colour than in the $3.6-8\micron$ colour used by Boyer et
al.   The latter  diagnostic  thus favoured  the detection  of more
dusty and cooler stars, while the $K-8\micron$ colour was more effective
in  distinguishing warmer  stars with  less dusty  envelopes.  Moreover,
 \citet{origlia10} examined  HST [Wide  Field Channel
(WFC) at the Advanced Camera for  Surveys (ACS)] images of each of 
their 78  dusty candidates to argue  that only three stars  had close and
relatively  bright   ($\pm\sim1$  magnitude)  companions   within  the
Spitzer/IRAC PSF area.

The  Boyer  group response  to  \citet{origlia10}  came recently  in
\citet{iana,ianb}. This time they added NIR photometry to the MIR data
analysed in \citet{boyer10}.
The   NIR   data   came   from   \citet[SOFI@ESO/NTT]{salaris07}   and
\citet[2MASS]{sku06}.  As  in Origlia et  al.  (2010), they  found that
CMDs using  $(K-8\micron)$ colours showed a red  sequence more clearly
than those using $(3.6\micron  - 8\micron)$ colours.  However, only 45
out of  the 93 sources  were confirmed as  IR excess stars in  the new
photometry, with {22 below $\sim1000L_{\odot}$}.
  This led  \citet{iana} to  suggest that  the  primary difference
  between the  two studies resulted  from differences in  the original
  photometric reduction methods.

\begin{figure}
\centering
\includegraphics[width=6.5cm]{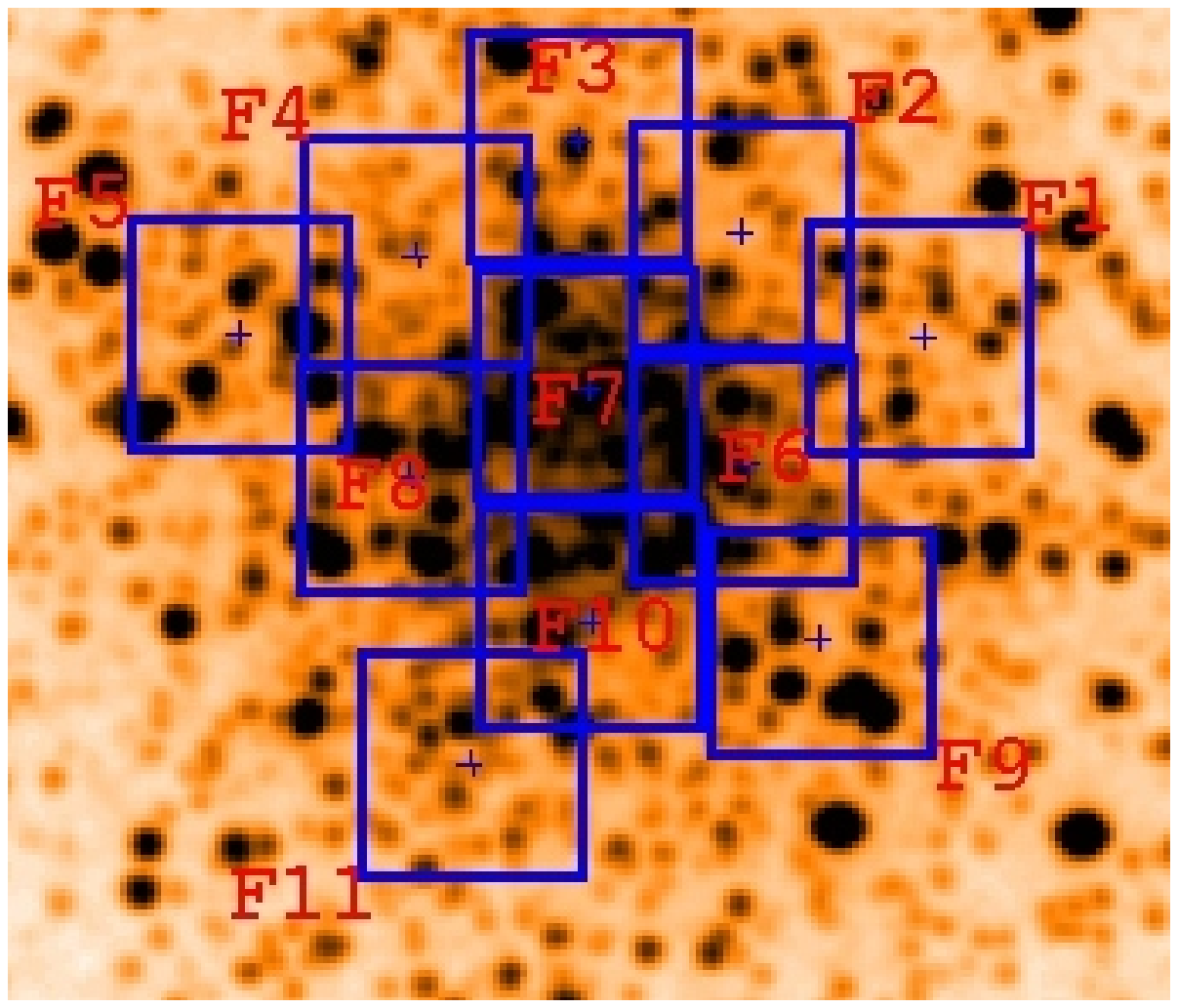}
\includegraphics[width=9cm]{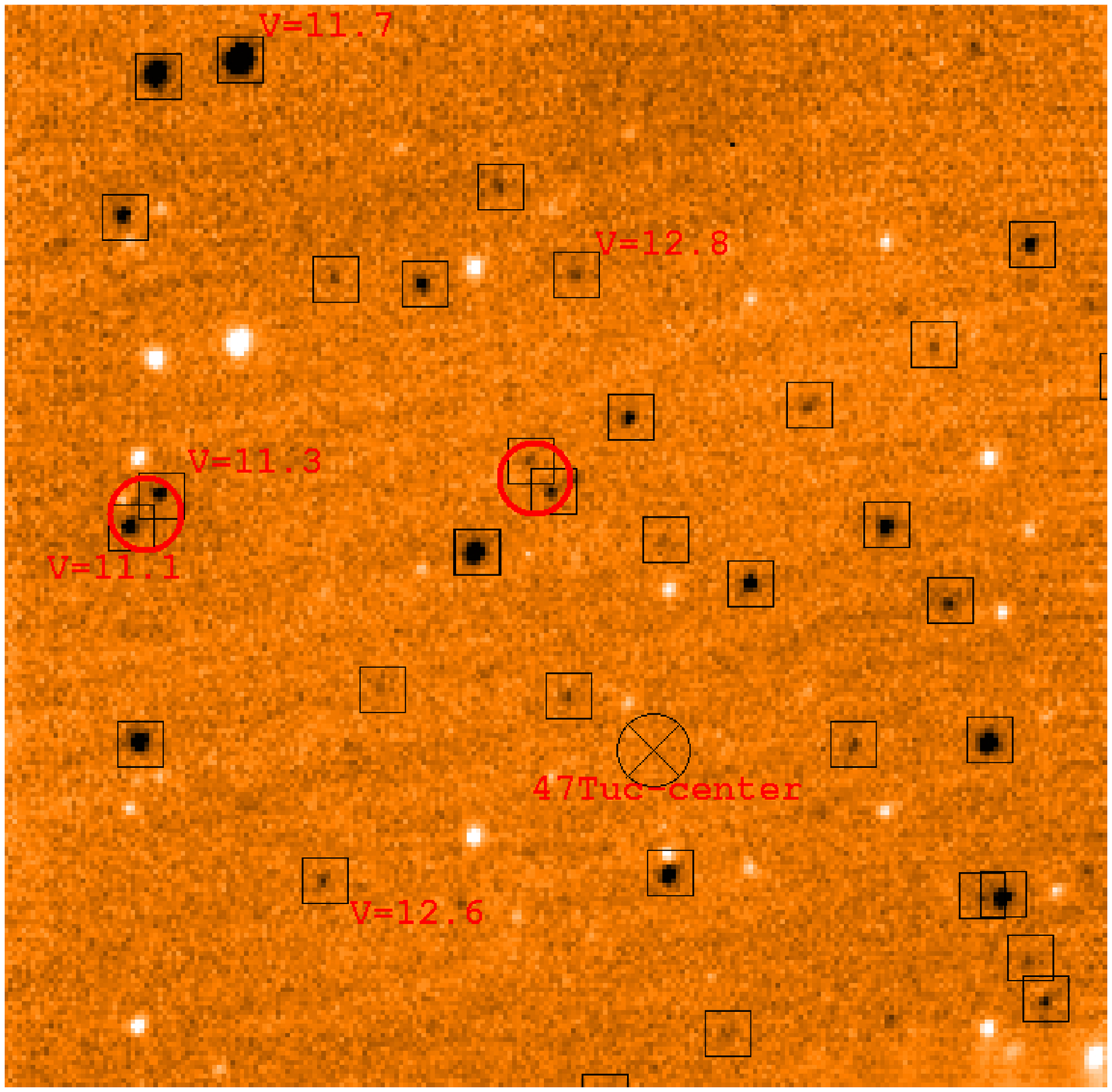}
\caption{Upper panel: the designed VISIR-Mosaic overlain on a
  2MASS $K-$band image. The lower panel shows the central (F\#7) VISIR
  $8.6\micron$ image. 
  Red  circles  with  $2\farcs0$  diameter  mark  areas  of  potential
  blending  caused by  lower-resolution  instruments.   The circled cross
  marks  the cluster  centre.  Open  squares highlight  the identified
  targets in pointing\#7.  Also  visible are some {\em negative beams}
  of stars  that are $8\farcs0$ vertically shifted.  This reflects the
  adopted chopping throw.}
\label{f_colour}
\end{figure}

Therefore, 47Tuc  still represents {\it the} key  cluster to establish
the luminosity  extent of  the mass-losing  giants. Ideally  one would
like to re-examine this  cluster with an appropriately high-resolution
mid-IR  instrument  that allows  a  proper  accounting  of the  higher
central crowding conditions.
Motivated by all the uncertainties still existing on the nature of the
mass loss  rate, we  decided to exploit  the superior  capabilities of
VISIR  to re-examine  the RGB  stars  of 47Tuc  with an  appropriately
high-resolution MIR instrument that  allows a proper accounting of the
higher central crowding conditions.
In  this paper  we  present 8.6$\micron$  ground-based imaging  data,
which indeed achieve $\sim8\times$ better spatial resolution than that
of Spitzer ($0\farcs3$~ vs. 2-3\arcsec).

In Sections \ref{s_data} and \ref{s_cal} we present the data reduction
and calibration,  in Sections  \ref{s_cmd} and \ref{s_K}  we construct
and  analyse the  combined optical-NIR-MIR  colour-magnitude diagrams.
Sections \ref{s_LPV} presents the identified long-period variables and
a spectroscopic study  of the brightest AGB star.  The conclusions are
summarized in Section \ref{s_conc}.

\begin{table*}
\caption{The journal of observations.}
\label{t_log}
\centering
\begin{tabular}{c c c r c c c r}
\hline\hline
Pointing & RA (J2000) & Dec. (J2000) & Offset                & \multicolumn{3}{c}{Obs.Date}   & Stars \\
         &            &              &       [$^{\prime\prime}$] & OB1       & OB2      & OB3     &       \\
\hline       
F1	& 00:23:55.9	& $-$72:04:36.9 &  48    &  2009$-$10$-$03 & 2009$-$10$-$03 & 2009$-$10$-$04  & 12\\
F2	& 00:24:01.7	& $-$72:04:21.4 &  37    &  2009$-$10$-$06 & 2009$-$10$-$06 & 2009$-$10$-$09  &  3\\
F3	& 00:24:06.9	& $-$72:04:08.1 &  45    &  2009$-$10$-$09 & 2009$-$09$-$09 & 2009$-$10$-$09  &  4\\
F4	& 00:24:12.0	& $-$72:04:24.6 &  40    &  2009$-$10$-$10 & 2009$-$10$-$10 & 2009$-$10$-$18  &  3\\
F5	& 00:24:17.7	& $-$72:04:36.3 &  57    &  2009$-$10$-$18 & 2009$-$10$-$22 & 2009$-$10$-$25  & 12\\
F6	& 00:24:01.5	& $-$72:04:55.4 &  20    &  2009$-$11$-$15 & 2009$-$11$-$15 & 2009$-$11$-$18  & 14\\  
F7	& 00:24:06.6	& $-$72:04:44.6 &   8    &  2009$-$11$-$13 & 2009$-$11$-$18 & 2009$-$12$-$04  & 32\\  
F8	& 00:24:12.2	& $-$72:04:57.3 &  30    &  2009$-$12$-$17 & 2009$-$12$-$17 & 2009$-$12$-$19  &  9\\  
F9	& 00:23:59.2	& $-$72:05:21.4 &  42    &  2009$-$12$-$24 & 2010$-$01$-$07 & 2010$-$01$-$29  &  7\\  
F10	& 00:24:06.4	& $-$72:05:18.9 &  27    &  2009$-$12$-$24 & 2010$-$01$-$26 & 2010$-$01$-$27  & 18\\ 
F11	& 00:24:10.3	& $-$72:05:40.1 &  53    &  2009$-$12$-$24 &  ---           & ---             &  5\\
\hline
\end{tabular}
\end{table*}

\section{Observations and data reduction}
\label{s_data}

\subsection{The VISIR imaging data set}

Our observations  were obtained using the VLT  spectrometer and imager
for  the  MIR (VISIR,  \citealt{lagage04})  currently  located at  the
Cassegrain focus  of UT3.  VISIR  provides diffraction-limited imaging
at high sensitivity in the two mid infrared (MIR) atmospheric windows:
the  $N$-band   between  $8-13\micron$   and  the  $Q$-band  between
$16-24\micron$.  The  observations were  obtained with the  VISIR PAH1
filter ($\lambda_{\rm  c}=8.59$, $\Delta~\lambda=0.42\micron$).  It is
designed to  avoid the telluric  ozone band and provides  a relatively
high sensitivity.

Our programme  [084.D-0721(A), P.I.  Momany] was  awarded 34~hours. We
used  the $0\farcs127$  pixel scale  for the  imager, which  yielded a
field-of-view of  $32\farcs5\times32\farcs5$. Three observation blocks
(OB) were dedicated to each of the 11 pointings, and each of these OBs
had  an effective  integration time  of 35  min., that  is a  total of
$\sim1.75$ hours per pointing.  These observations were carried out in
service mode in the period between October $3$, 2009 and January $30$,
2010.
Table~\ref{t_log} reports the J2000 right ascension and declination of
the  11   VISIR  pointings.   Also   reported  are  the   offsets  (in
arc-seconds) from the cluster centre  (as in Goldsbury et al. 2010) of
each  pointing and  the dates  at which  the OBs  were  observed. Last
column reports the number of stars found in each pointing.

As  a result, the  observations were  spread over  a four  month period,
obtained under various weather  conditions. As for any ground-based mid-IR
observations, the technique of chopping  and nodding was used to
enhance  the detection of  target stars  over the  background emission
originating both from  the sky and the telescope.   The declination of
47Tuc  does not allow  airmasses below  1.4 and  around $90\%$  of the
obtained  images had  airmass between  1.4  and 1.8.  The varying  MIR
weather  conditions  (i.e.   water-vapor  content,  temperature,  etc)
impacted heavily  on the delivered quality  of the data.  In turn, the
combination of  the three  OBs (of any  given pointing) was  not possible,
even when these OBs were separated by only one day.
Therefore we  performed the photometric reduction  (i.e.  detection of
point-sources  and derivation  of  their aperture  magnitudes) on  the
reduced image of each single OB {\em separately}.

The  tested  DAOPHOT  (\citealt{stetson87})  package was  employed  to
perform the photometry on the reduced image for each  OB.
The {\sc FIND} routine was set to a detection threshold of three times
the  background standard  deviation  estimated on  a  given image.  In
general, {\sc  FIND} detected about two  to three times  the number of
the stars  that one could identify  by eye.  A  careful examination of
the {\it  extra} detections revealed  that these are caused  by abrupt
fluctuations in the estimated background level or hot pixels. 
Indeed,  even though we employed  a  relatively  relaxed
constraint, wherein a detection was  considered {\em real} if found in
at least two images, these  {\it extra} detections did {\em not} survive
the {\sc DAOMASTER} matching process for a given field.

A powerful spurious-detections filtering process was performed
when catalogues from at least two filters were combined.
In this  framework we  decided to employ  the astrometric/photometric
ACS@HST $m_{\mathrm {F606W}}$,$m_{\mathrm {F814W}}$ catalogue of 47Tuc
(\citealt[kindly  provided by the  author]{jay09}).  This catalogue
provided two advantages.  The first  was that the optical HST data were
$\sim100\%$ unaffected by  photometric {\it incompleteness} around our
VISIR  data sampling.   Indeed,  the CMDs  presented in  \citet{jay09}
sampled $\sim8$  $m_{\mathrm {F606W}}$  magnitudes below the  RGB tip,
whereas  our  VISIR  data  address only  the  brightest  $\sim2.0-2.5$
magnitudes.  Moreover, the catalogue of  Anderson et al.  was based on
the  relatively  {\em  red}  F606W  and  F814W  filters,  i.e.   still
sensitive to red sources with MIR-excess.
  The second advantage was that the ACS spatial resolution (pixel size
  of  $0\farcs05$)  is very  close  to that  of  our  VISIR data.   We
  therefore  confidently   used  this  HST  catalogue   in  the  VISIR
  point-source  detection process.  In other  words, the  ACS absolute
  astrometric   positions  were   transformed  to   VISIR   X,Y  pixel
  coordinates, and a subset of this catalogue ($V\le13.0$) was assumed
  to provide the detected point-sources.

  In  general, all  objects  {\em visible}  in  the VISIR  $N_{\mathrm
    {8.6\micron}}$  images  had   an  optical  HST  counterpart  (with
  $V\le13.0$).
This  confirms that the  use of the Anderson  et al. HST
  catalogue did  not introduce  blue-filters selection effects  in our
  final VISIR/ACS catalogue.
There was one interesting case for  which a visible VISIR target had a
fainter optical counterpart,  see Appendix~A.
Thus, this combined ACS/VISIR  detection strategy provided us with the
best possible elimination of spurious-detections.

Secondly,the respective  catalogues were  merged for each  pointing to
produce a single aperture magnitude catalogue.
  These  magnitudes  (see   Table~2)  were combined using   the
  DAOPHOT/DAOMASTER     task,    which         computed    an
  intensity-weighted mean of the  three aperture magnitudes in each  field.

The  upper panel  in Fig.~\ref{f_colour}  displays the  2MASS $K$-band
view of  the core  of 47Tuc.   Overlaid are the  footprints of  the 11
VISIR $32\farcs5~\times~32\farcs5$  pointings. Given the  service mode
nature of the programme and the necessity to ensure a common photometric
scale for  the single fields,  the pointings were designed  to overlap
on bright  stars ($K\le10$ magnitude). 
Typically 2-4 such bright stars  were shared among adjacent fields.  A
relative zero-point  was computed from  their magnitudes to  bring all
pointings on  the photometric scale  of pointing  \#7, which  covered the
cluster core.
The  lower panel  of Fig.~\ref{f_colour}  displays the  VISIR pipeline
reduced  image of a  {\em single}  35-min OB.   There is  a one-to-one
correspondence   between  all   $m_{\mathrm {F606W}}\le13.0$   stars  from   the
\citet{jay09} HST catalogue and VISIR detections.
Although limited to  a few cases, close stars (e.g.  the two examples of
stars falling within a $2\farcs0$ diameter circle) end up as blends in
lower-resolution  instruments. Lastly, adopting  the HST determination
of the cluster centre derived  by \citet{gold10}, it is clear that our
VISIR $8.6\micron$  imaging data covers  the inner and  most crowded
$1\farcm15$ core of the cluster.

\subsection{VISIR spectroscopic data set}
The VISIR  $N$-band spectroscopy [Prog.60.A-9800(I)]  consisted of two
low-resolution  ($R\sim300$)   setups  with  central   wavelengths  at
$9.8\micron$  and  $11.4\micron$ obtained  with  the $1\farcs0$  slit.
These two setups were chosen  to allow the coverage of the $10\micron$
silicate  feature   in  a  particularly  bright  AGB   star  (V8,  see
Section~\ref{s_LPV}),  which  we observed  to  assess  the quality  of
VISIR's  spectroscopic performance.  To  properly calibrate  the data,
the Standard VISIR Telluric HD4815 was observed using the same setup.
The data  reduction  was  performed in
two steps.  First,  the ESO/VISIR pipeline was used  to extract the 2D
frames of both the science target and the telluric.
Later,  we used the  pipeline developed  by \citet{honig10},  which is
specifically  designed  to  perform  a  proper  background  correction
(including  sky and  periodic detector  signals), beam  extraction, 2D
flux calibration and 1D spectrum extraction.
The 2D spectrum of V8 and  its calibrator HD4810 showed very small FWHM
variations. In particular, the  $9.8$ and $11.4\micron$ gratings of V8
showed a FWHM of  $\sim 0\farcs40$ and $\sim 0\farcs45$, respectively.
To  recover  $\sim 96\%$  of  the flux  in  both setups,  we
employed  an  extraction  aperture   of  $\sim  0\farcs75$  and  $\sim
0\farcs80$  for the  $9.8$ and  $11.4\micron$  gratings, respectively.
Lastly, we note that the two ($9.8$ and $11.4\micron$) reduced spectrum
of V8 showed almost perfect  flux matching in the overlap wavelength
region.

\subsection{The HST catalogue}

As  explained above,  the astrometric/photometric  ACS@HST $m_{\mathrm
  {F606W}}$   and    $m_{\mathrm   {F814W}}$   catalogue    of   47Tuc
(\citealt{jay09}) was  used as  a reference catalogue  for identifying
the point sources in VISIR's  images.  We refer the reader to Anderson
et  al.'s  paper  for   details  concerning  the  data  reduction  and
calibrations.   We complemented  the Anderson  et al.   catalogue with
independent   measurements  in   $m_{\mathrm   {F336W}}$,  $m_{\mathrm
  {F435W}}$,  $m_{\mathrm {F606W}}$  and $m_{\mathrm  {F814W}}$.  This
extends  our final ($m_{\mathrm  {F606W}}$, $m_{\mathrm  {F814W}}$ and
$N_{\mathrm {8.6\micron}}$) catalogue  with blue $m_{\mathrm {F336W}}$
and $m_{\mathrm  {F435W}}$ measurements to avoid  selection effects on
high mass loss rate stars.
The latter  two photometric bands  are based on reduction  of ACS/Wide
Field Channel (WFC) images  from programme GO-10775, while $m_{\mathrm
  {F435W}}$ is  based on  a single 10-second  image from  GO-9281.  In
both cases  the photometric reduction  and calibration of  the ACS/WFC
data was  carried out  using the software  presented and  described in
detail in  \citet{jay08}. The $m_{\mathrm  {F336W}}$ measurements were
based on a 30-second exposure from GO-11729 obtained with the HST Wide
Field Camera~3  (WFC3). This image was pre-reduced  using the standard
HST pipeline, while the fluxes were measured using a software based on
the \textit{img2xym-WFI} (\citealt{jay06}) that will be presented in a
separate paper (Anderson et al. in prep.).

\subsection{The near-infrared catalogue}
\label{s_satur}
To sample  the core  of 47Tuc  in the NIR,  we searched  for available
catalogues and archival ESO  data.  Unfortunately, given the extremely
crowded conditions  of the  47Tuc core, observers  tend to  avoid this
region.
We used the excellent NIR  $JHK_{\rm s}$  catalogue of  47Tuc   of
\citet{salaris07}.  This  was based  on  high-resolution ESO/NTT  SOFI
infrared images (pixel size  of $0\farcs29$) obtained under excellent
seeing conditions $\le0\farcs9$.
However, the catalogue of Salaris  et al.  does not fully encompass the
cluster  core, but only  the northern  half.  Moreover,  the excellent
seeing conditions resulted in saturation of the 47Tuc upper RGB.

An archival  SOFI@NTT [69.D-0604(A)] $JHK_{\rm s}$ data set of  47Tuc was also
analysed.   The data set  was reduced  following the  standard methods
presented in  \citet{momany03}.  Unfortunately, this  data set covered
mainly  the southern  part around  the  cluster core,  with a  limited
$30\farcs0$ overlap region with the Salaris et al.  catalogue.
The  seeing  on  the  averaged  $J$- and  $K_{\rm s}$-band  was  $1\farcs4$  and
$1\farcs3$,  respectively, significantly  worse than  the data  set of
\citet{salaris07}.  The resulting NIR diagrams still showed saturation
in  the upper  RGB  and higher  dispersion  along the  RGB.  We  
therefore mainly used the \citet{salaris07} catalogue.

\section{Calibration of the VISIR data}
\label{s_cal}

With the  mosaicing strategy  described above, the  merged photometric
catalogue  of the 11  pointings was  put on  the photometric  scale of
pointing\#7. The  zero-point offsets were estimated from  the stars in
the  overlapping regions.  In  this manner,  the airmass  and detector
integration time (DIT) of pointing  \#7 could be used to calibrate the
entire merged catalogue.
We use the standard star  HD4815-K5III, observed just before the first
OB of pointing \#7, to flux-calibrate our data.
The airmass of  the HD4815 was $1.569$, while that  of pointing \#7 was
$1.476$.
Although the calibration  process is straightforward, ground-based MIR
calibration   can  be   tricky;  we   therefore  provide   a  detailed
presentation.   First of  all,  we  relied on  the  VISIR pipeline  to
deliver  reduced images  normalized to  1-second exposures.   The {\em
  instrumental} magnitude of the  standard star, which was observed in
perpendicular  CHOPPING/NODDING   mode,  was  measured   by  the  {\sc
  DAOPHOT/PHOT}  task.   Four  measurements  were  obtained,  i.e.  two 
measurements  from  each of  the  positive  and  negative beams.   The
averaged  four measurements provided  a standard  deviation of  less than
$\sim2\%$. Before converting this average magnitude into a flux value,
there remained  two corrections to  be applied: the  aperture correction
and the extinction.

\subsection{Aperture correction}

The  first  correction   applied to  the  instrumental
  magnitudes  is to correct for the loss  of flux  outside  the adopted
  aperture radius: the so-called {\it aperture-correction}.
This is  usually made by  estimating the magnitude  difference between
the  employed aperture  radius ($2.5$-pixels,  $0\farcs32$)  and those
obtained  at  larger radii  (out  to  $16$-pixels, $2\farcs03$).   The
aperture correction is measured  at radii corresponding to a near
constant  magnitude difference,  i.e.   a plateau.  In  our case,  the
standard star aperture correction was quite significant: around $0.37$
magnitude.

\begin{figure}
\centering
\includegraphics[width=7cm]{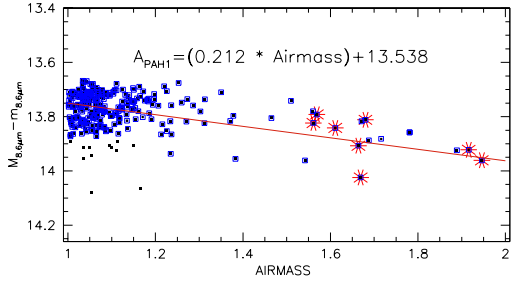}
\caption{Estimating  the extinction  coefficient for  the $0\farcs127$
  PAH1 filter  at VISIR.   Multiple  observations of the  standard star
  HD4815 are highlighted with red symbols. }
\label{f_ext}
\end{figure}

\subsection{Extinction coefficient}

The second correction is that  related to the airmass of the 47Tuc
  VISIR  observations.  Given  the cluster  declination  (and implied
high  airmass  of $\ge1.4$),  the  extinction  could  not be  ignored.
Moreover,   the   extinction    coefficient   of   the   PAH1   filter
($\lambda_{\rm c}=8.59\micron$) has not been measured before.

For  this purpose,  we made  use of  an  ESO-archival page\footnote{
\textsf{http://archive.eso.org/bin/qc1$_{-}$cgi?action=qc1$_{-}$browse$_{-}$table\&} \\ 
\textsf{table=visir$_{-}$zp$_{-}$img}}
to collect PAH1 standard star measurements for the past six years.
Availing ourselves of the  measured instrumental flux, and knowing the
standard's true flux (in Jansky, hereafter Jy), we show the results in
Fig.~\ref{f_ext}. Only  PAH1 with the  $0\farcs127$-setup measurements
are  reported, and  a $2.5\sigma$  clipping  has been  applied to  the
data-points  for bins  of airmass  of $0.1$.  
A  least-square  fit   to  $\sigma$-clipped  data-points  provided  an
extinction coefficient of $0.212$  mag.  This may seem relatively high
(compared  for example  to  the $0.05$  mag.   for $K_{\rm  s}$-band),
however, although  there is a  scarcity of published  ground-based MIR
extinction values, $0.212$  was found to   perfectly agree  with
technical
tests\footnote{\textsf{http://www.eso.org/sci/facilities/lasilla/instruments/timmi/Reports/}
  \textsf{oschuetz/Projects/T2$_{-}$Extinc/TIMMI2$_{-}$extinc.html\#3.2}}
made in a similar  manner on  TIMMI2 (previously mounted  on the
ESO/La Silla 3.6m telescope).

In the  specific  case of  HD4815,  the instrumental  magnitude
($m_{\mathrm {inst.}}$) corrected  for airmass and aperture correction
({\it apert. corr}) corresponds to:

\begin{equation}
  m_{\rm {inst.}}\,=\, m\,-(apert.corr.)\,-(ext.coef.\times\, airmass)
\end{equation}
 \begin{equation}
m_{\rm {inst.}}=m\,-(0.370)\,-(0.212\times\,1.569)
\end{equation}

Hence, the two corrections  account for up to
$\sim0.7$ mag.,  making our photometric scale brighter.

\subsection{Absolute fluxes and magnitudes}

After  correcting the  standard  star instrumental  magnitude, we 
converted this into  a flux value. This instrumental  flux was divided by
the  standard  star's known  flux  (in  Jansky  units) to provide  the
conversion factor (ADU/Jy).

To derive the absolute calibration of the 47Tuc PAH1 photometric
catalogue, we  first applied the  aperture and extinction  corrections (as
 performed  for  the standard  star).  Next,  these corrected  instrumental
magnitudes were converted into fluxes and each star was divided by the
conversion factor,  obtained from the standard star. This basically
converted  our instrumental  magnitudes  into absolute  flux values  in
Jy.
The absolute flux  ($F_{\nu}$,  in units of Jansky) could then be
converted  into  magnitudes on  the  (i)  {\em AB}  (\citealt{oke83})
photometric system:

\begin{equation}
m_{\rm {AB}}\,=\,-\,2.5\, \log\,\Big({F_{\nu}}\times{10^{-23}}\Big)\,-\,48.57,
\end{equation}

and  (ii) the Merlin-N1 photometric system:

\begin{equation}
m_{{\rm N1}}\,=\,-\,2.5\, \log\,\Big(\frac{F_{\nu}}{49.4}\Big).
\end{equation}

The zero-point magnitude for the Merlin-N1 photometric scale ($49.4$%
\footnote{\textsf{http://ssc.spitzer.caltech.edu/warmmission/propkit/pet/magtojy/} \\
  \textsf{index.html\#fnu$_{-}$to$_{-}$mag}}  Jy) refers
to an effective wavelength of $8.81\micron$. This is slightly redder than
the       employed        PAH1       filter       ($\lambda_{\rm c}=8.59$,
$\Delta~\lambda=0.42\micron$).   However,  we note  that  there are  no
other $\sim8\micron$  filters in the UKIRT  or in  the Johnson
systems (the closest being $\sim10.1\micron$).

We caution that the AB and Merlin-N1 system differ systematically by a
zero-point of  $4.696$ magnitude. For the  remainder of this  paper we will
use  the Merlin-N1 magnitude  scale, and  for brevity,  refer to  it as
$N_{\mathrm {8.6\micron}}$.
%

\begin{figure}
\centering
\includegraphics[width=9cm]{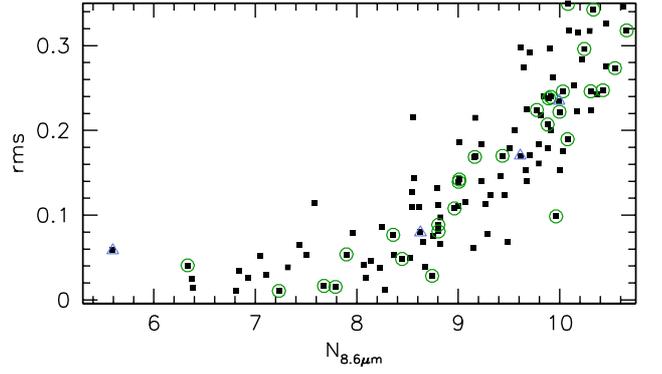}
\caption{Distribution  of  standard deviations  (of  stars with  three
  photometric   measurements)  as  a   function  of   the  $N_{\mathrm
    {8.6\micron}}$  magnitude.   Open  circle  highlight   stars  from
  pointing\#7, while open triangles highlight those from pointing\#3.}
\label{f_rms}
\end{figure}

\subsection{Photometric errors}

  Ideally, photometric errors should be estimated from artificial star
  experiments   by  comparing   the   input/output  magnitudes   (e.g.
  \citealt{momany08}).  However, for the majority of the 11 pointings,
  there  were not enough detected  stars to  allow the  construction of
  reliable   point   spread   functions   (PSF)  necessary   for   the
  simulation/injection of artificial stars.
  On   the  other   hand,  each   pointing  had   a  total   of  three
  OBs\footnote{The  VISIR  service  mode  observations  adopt  a
      reference flux  level (expressed in  terms of mJy/hour)  that is
      measured on  MIR standard stars observed either  before or after
      the science  OB.  There were  few OBs (of  the order of  1-2 per
      pointing) that exceeded the  reference flux values and these OBs
      were excluded  from our analysis.}  and  the standard deviation
  of  these three  measurements  were used  to  trace the  photometric
  errors in a given pointing.
  Figure~\ref{f_rms}  displays   the  distribution  of   the  standard
  deviation as  a function of  the $N_{\mathrm {8.6\micron}}$ magnitude  for stars
  with three  measurements.  The  distribution shows  the expected
  increase of the  rms as a function of  magnitude.  In particular, no
  significant   rms    variations   were   found    among   different
  pointings.   The   average   photometric   errors  are   listed   in
  Table~\ref{t_errors}.

\begin{table}
\caption{Average $N_{\mathrm {8.6\micron}}$ photometric errors  estimated 
from the rms distribution in Fig.~\ref{f_rms}.}
\label{t_errors}
\centering
\begin{tabular}{c c | c c}
\hline\hline
$N_{\mathrm {8.6\micron}}$ & rms & $N_{\mathrm {8.6\micron}}$ & rms \\
\hline       
    6.000  &  0.027  &   8.500  &  0.083 \\
    6.500  &  0.027  &   9.000  &  0.114 \\
    7.000  &  0.033  &   9.500  &  0.151 \\
    7.500  &  0.045  &   10.000 &  0.232 \\
    8.000  &  0.058  &   10.500 &  0.332 \\
\hline
\end{tabular}
\end{table}

In  addition   to  the  photometric  internal   errors  quantified  in
Fig.\ref{f_rms}, we lso have to consider the one caused by shifting of
the photometric  scales of  the 10 pointings  with respect to  that of
pointing\#7.  We quantified this  additional error by first collecting
the residuals around  the average shift (as derived  from the stars in
the  overlapping  region of  two  adjacent  fields), and  subsequently
estimating the rms  of all the residuals. This turned out  to be of the
order of $\sim0.015$ magnitude.

  Lastly, we  estimated the  systematic uncertainty in  calibrating the
  $N_{\mathrm {8.6\micron}}$  data.   The   sources  of  uncertainty  are  those
  associated with the aperture-correction method and that from 
  the derivation of the extinction coefficient in $N_{\mathrm {8.6\micron}}$.
  The aperture-correction uncertainty is of the order of $0.031$ mag.,
  estimated from  the consistency of  the value derived for  the three
  single images of pointing\#7.
  The   uncertainty  associated   with   the  extinction   coefficient
  derivation is of the order  of $0.033$ mag., estimated as the error 
  on the slope of a least-square fit (as in Fig.~\ref{f_ext}).
  Overall,  the  total  zero-point  uncertainty,  obtained  from  the
  quadratic sum of the above two factors is $\sim0.045$ magnitude.

  In   Table~\ref{t_xonline}   we   report   the   VISIR   $N_{\mathrm
    {8.6\micron}}$ catalogue of  the 47Tuc central $1\farcm15$ region,
  along  with  the  $m_{\mathrm  {F606W}}$ and  $m_{\mathrm  {F814W}}$
  magnitudes (from  Anderson et al.  2009),  the $m_{\mathrm {F336W}}$
  and $m_{\mathrm  {F435W}}$ magnitudes  (reduced in this  paper), the
  $J$- and $K$-magnitudes (from Salaris  et al.  2007), the flux value
  in Jansky and the known long-period variables (LPV) identification.

\begin{figure}
\centering
\includegraphics[width=9cm]{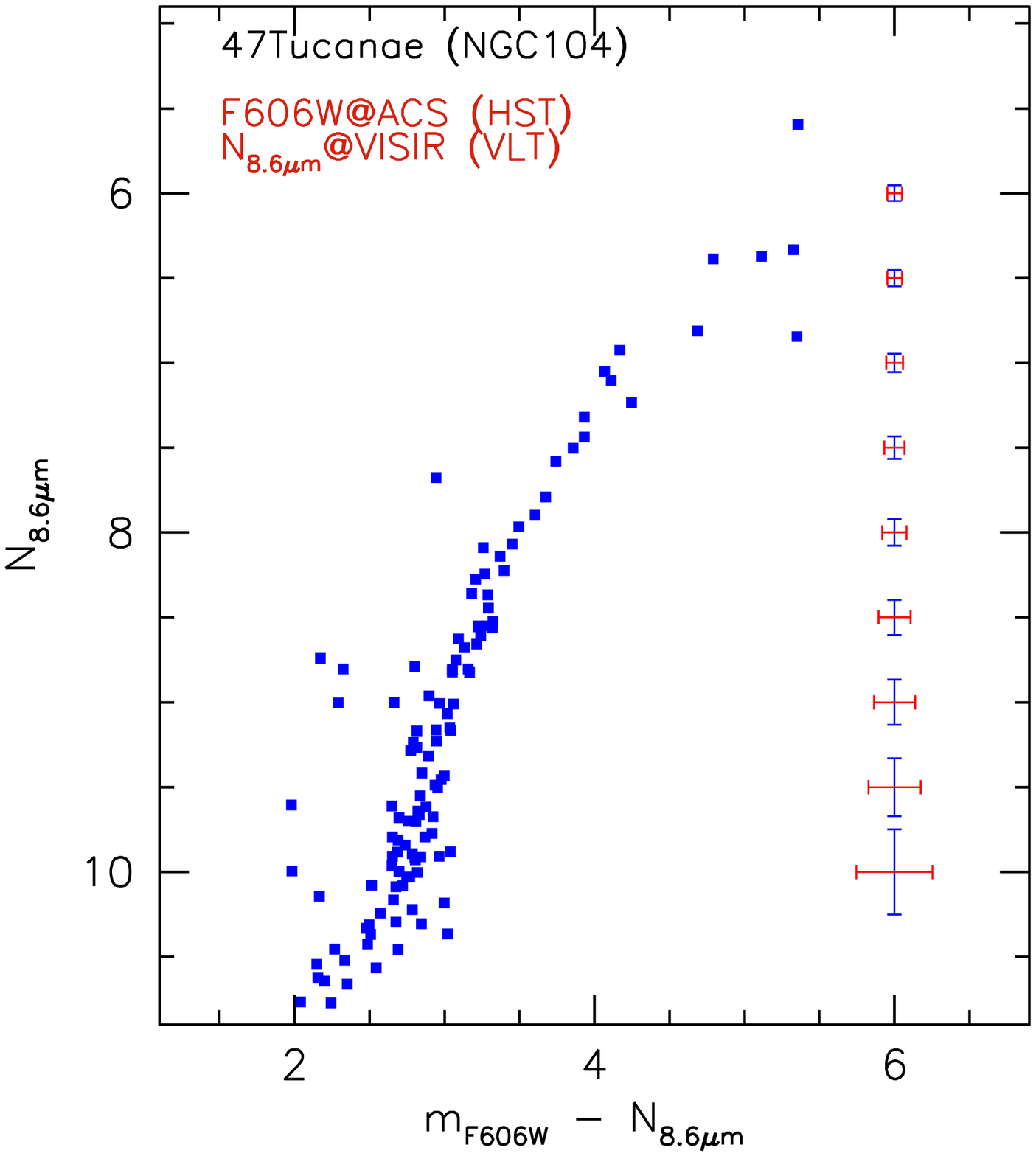}
\caption{Combined        ACS@HST       and        VISIR@UT3
  $N_{\mathrm {8.6\micron}}$,($m_{\mathrm {F606W}}-N_{\mathrm {8.6\micron}}$)   CMD of 47Tuc. }
\label{f_cmd}
\end{figure}

\section{HST/VLT colour-magnitude diagrams}
\label{s_cmd}

\begin{figure}[h]
\centering
\includegraphics[width=8.cm]{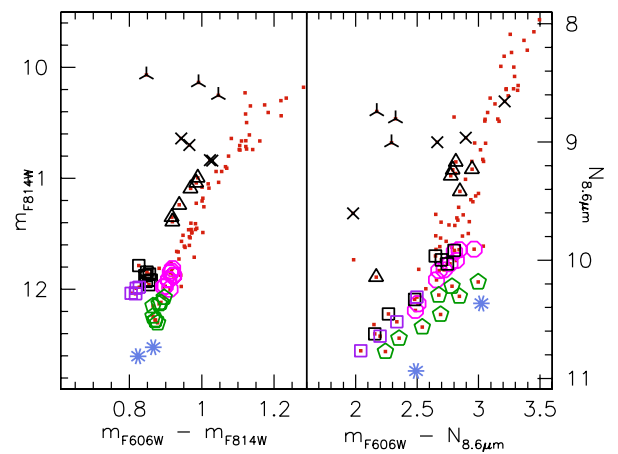}
\caption{Identification of AGB stars selected from the optical
  HST CMD  on the combined  $N_{\mathrm {8.6\micron}}$,($m_{\mathrm {F606W}}-N_{\mathrm {8.6\micron}}$)
  diagram of 47Tuc. The symbols in both panels trace particular (early
  and late)  AGB groups  as well  as an arbitrary  selection of  the three
  faintest RGB groups.}
\label{f_faintest}
\end{figure}

The          combined          ACS@HST          and          VISIR@VLT
$N_{\mathrm {8.6\micron}}$,($m_{\mathrm {F606W}}-N_{\mathrm {8.6\micron}}$) colour-magnitude diagram
  of 47Tuc core  is presented  in Fig.~\ref{f_cmd}.   The plotted
error  bars in  both magnitude  and colour  are mainly  driven  by the
photometric  uncertainty  in  $N_{\mathrm {8.6\micron}}$,  while  those   in  the  HST
$m_{\mathrm {F606W}}$ filter  are lower  than $0.002$ magnitude.   The brightest
star  at  $N_{\mathrm {8.6\micron}}=5.595$  is  a  known  variable star  (V8,  see
Sect.\ref{s_LPV}) with  a total flux of $0.286$~Jy,  while the faintest
reliable  measurements reached (see  Fig.~\ref{f_faintest}) are
for stars at $N_{\mathrm {8.6\micron}}\simeq10.0$ ($\sim0.005$~Jy).

The  CMD  shows  that  we  sampled  the  upper  $\sim4$  magnitudes  in
$N_{\mathrm {8.6\micron}}$  ($\sim3$ magnitudes in $m_{\mathrm {F814W}}$ filter)  of the red
giant branch  of 47Tuc.   The horizontal branch  level is  expected at
$N_{\mathrm {8.6\micron}}\simeq12.1$, that is $\sim1.7$ magnitude fainter than our
photometric  incompleteness level  at  around $N_{\mathrm {8.6\micron}}\sim10.4$.
Clearly, the  VISIR ground-based observations did not  compete with the
deeper Spitzer CMDs. 
%

The smoothly  curved morphology  of the upper  RGB in  the $N_{\mathrm
  {8.6\micron}}$,($m_{\mathrm    {F606W}}-N_{\mathrm   {8.6\micron}}$)
plane  resembles that  of  optical $m_{\mathrm  {F814W}}$,($m_{\mathrm
  {F606W}}-m_{\mathrm  {F814W}}$) diagrams  published  in Anderson  et
al. (2009).
The    $N_{\mathrm   {8.6\micron}}$,($m_{\mathrm   {F606W}}-N_{\mathrm
  {8.6\micron}}$) CMD is not ideal to check for MIR-excess giants (see
Fig.\ref{f_tau1}). Still, we note that there are no clear ``outliers''
(RGB or AGB) from the cluster mean RGB locus.
  Indeed,   increasing    photometric   errors   around   fainter
  $N_{\mathrm {8.6\micron}}$ magnitudes can confuse any RGB/AGB separation.
  To investigate  this problem, we  identified all AGB  candidates using
  the  optical $m_{\mathrm  {F814W}}$,($m_{\mathrm {F606W}}-m_{\mathrm
    {F814W}}$)  HST  diagram  and   examined  their  position  on  the
  optical/MIR diagram (see Fig.~\ref{f_faintest}).
  Thanks  to  the HST  photometric  accuracy,  AGB  stars had  a  good
  ($m_{\mathrm  {F606W}}-m_{\mathrm {F814W}}$)  colour-separation from
  equally bright  red giants and the  group of stars  belonging to the
  so-called  early-AGB  (at  $m_{\mathrm {F814W}}\sim11.9$)  could  be
  easily separated from RGB stars.
Figure~\ref{f_faintest} shows  that the selected AGB  group, at most,
overlap  the RGB  mean locus.   In particular,  for stars  fainter than
$N_{\mathrm {8.6\micron}}\simeq10.3-10.4$,    the    photometric uncertainties   and
incompleteness appear  strong enough to cause a  sudden break in
the  slope of  the RGB.   This  is clearly  not a  real feature,
 hence stars fainter than $N_{\mathrm {8.6\micron}}\simeq10.4$ should not
be considered.

\begin{figure}
\centering
\includegraphics[width=9cm]{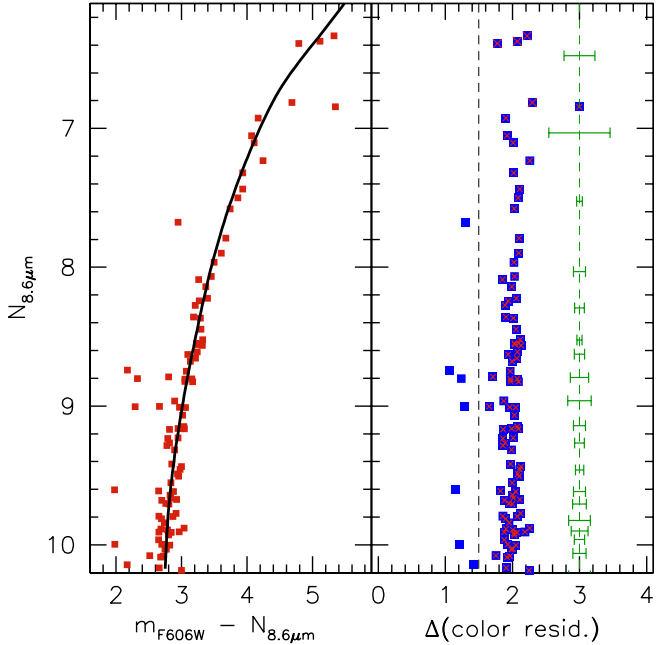}
\caption{Estimating the width of the RGB. The left-hand panel displays
  the    derived   RGB    mean   loci.    Stars    with   ($m_{\mathrm
    {F606W}}-N_{\mathrm {8.6\micron}}$)$\le1.5$  were excluded in this
  count. The right-hand panel  displays the straightened RGB sequence,
  along which a  bin of fixed-number of stars (5)  is used to estimate
  the mean colour dispersion. For clarity, these are shifted to a mean
  colour of $3.0$.}
\label{f_fig3}
\end{figure}

\subsection{The intrinsic width of the RGB}
\label{s_rgb} 
Related to the MIR-excess issue is the RGB intrinsic width, which
apparently remains small all over the sampled $\sim3$ $N_{\mathrm
  {8.6\micron}}$ magnitudes.
A quantitative estimate of the  RGB width is obtained by straightening the
RGB  (with respect  to a  fiducial line  or an  isochrone),  and, say,
excluding any  $2.5\sigma$ outliers  (within a specified  magnitude bin
size). This is presented  in Fig.~\ref{f_fig3}.  
Excluding the  upper two bins (relative  to the AGB  stars brighter than
RGB tip)  one appreciates how the  thickness of the  RGB is relatively
uniform over $\sim3$ magnitudes along  the RGB.  The average RGB width
turns out to be $0.095\pm0.038$, and no MIR-excess stars are detected.
One  has to  apply  a very  strong  $1.0\sigma$ clipping   to
identify outliers, and when this is applied, we note that the outliers
are found  mostly on the  blue side of  the RGB, therefore they are AGB
candidates.

\subsection{Isochrone fitting}
\label{s_isoc}

\begin{figure}
\centering
\includegraphics[width=8.5cm]{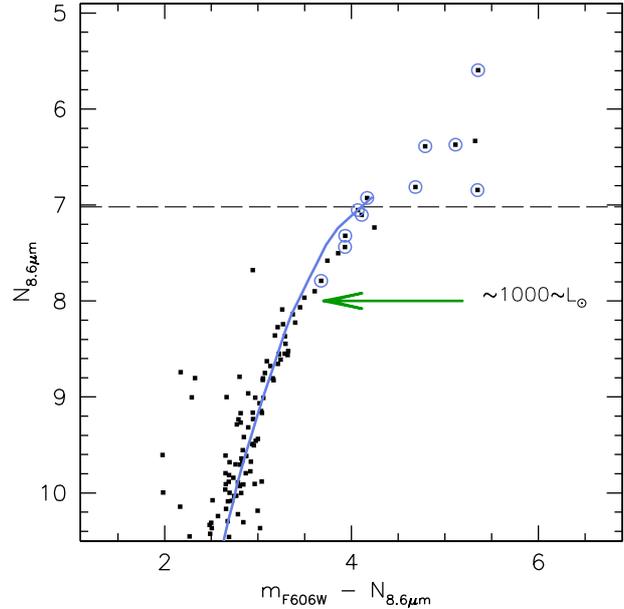}
\caption{Optical-MIR diagram  along  with an  isochrone from  the
  Padova library (\citealt{marigo08}).  Adopted  are an age of $12.59$
  Gyr, metallicity  of $Z=0.004$, distance  of ($m-M$)$_{\circ}=13.38$
  and reddening  of $E_{\mathrm {B-V}}=0.055$.  The dashed  line marks the  RGB tip,
    while   the   horizontal   branch  level   is   expected   at
    $N_{\mathrm {8.6\micron}}\simeq12.1$.  Open circles highlight the verified LPVs.
    The green arrow marks the  $\sim1000L_{\odot}$ luminosity level.}
\label{f_isoc}
\end{figure}

The  isochrone  fitting  is  usually employed  to  derive/confirm  the
studied cluster distance,  reddening and age. Our shallow  data do not
allow this level of analysis. Nevertheless, it  provides an excellent
test  for  ground-based calibration  and,  as we  shall see,  sheds
lights on an interesting aspect of the RGB morphology. For the cluster
distance and reddening we adopted  the values derived in 
\citet{eug00}, namely ($m-M$)$_{\rm V}=13.55$ and $E_{\mathrm {B-V}}=0.055$.
We  adopted  $A_{\rm V}=3.1\times~E_{\mathrm {B-V}}$, $A_{\mathrm {F606W}}=2.809\times~E_{\mathrm {B-V}}$,
and a null $A_{N_{\mathrm {8.6\micron}}}$.
In  Fig.~\ref{f_isoc}  we  display  the optical-MIR  diagram  with  a
theoretical      isochrone      from      the      Padova      library
(\citealt{marigo08})\footnote{\textsf{The isochrones  are available at
    http://stev.oapd.inaf.it/cmd}}.   The  isochrone was  ``coloured''
with the  $N_{\mathrm {8.6\micron}}$ Vega magnitude using the  VISIR PAH1 filter
transmission  curve   (see  \citealt{girardi02}  for   details).   The
presented  isochrone   has  an  age  of  $12.59$   Gyr  and  $Z=0.004$
([Fe/H]$=-0.67$ as derived  in \citet{eug00} and \citet{alves05}.
Overall,  the isochrone excellently  agree  with the  RGB
morphology.
%

For    giants   brighter   than
$N_{\mathrm {8.6\micron}}\simeq8.0$   the  observed  RGB   morphology  seemingly
deviate to redder colours than the isochrone loci.
Interestingly, this magnitude level corresponds to luminosities of the
order of $\sim1000L_{\odot}$, {\em suggested by \citet{iana} to indicate
  the beginning of dusty mass loss for giants}.
 This  perfectly   agrees  with  the  AKARI   study  by
  \citet{ita07}, who concluded that dust emission is mainly detected in
  AGB  variables, although there  are some  variable stars  with dust
  emission located below the RGB tip.
  Although this  is very  interesting, we note  that we  cannot firmly
  establish a  connection between this small colour  deviation and the
  onset of the dusty mass loss for giants below the RGB tip.
  Indeed,  the stellar  models of  this   cool $T_{\rm eff}$  can be  significantly
  affected  by small  errors in  opacities  and in  the mixing  length
  treatment.   Moreover, one  should  keep in  mind  that the  Kurucz
  spectra that were  used  to estimate  the  colour  transformations may  have
  additional sources of errors at this cool $T_{\rm eff}$.
  Consequently,  the above  factors caution  against  emphasizing this
  systematic deviation between the data and the isochrone.

  In conclusion, the isochrone fitting to both the observed RGB colour
  and  RGB-tip  magnitude-level  indicates  an  excellent  theoretical
  handle of  the unusual VISIR photometric  system and well-calibrated
  data.

\section{The $K_{\rm s}$, ($K_{\rm s}-N_{\mathrm {8.6\micron}}$) diagram}
\label{s_K}
\begin{figure}
\centering
\includegraphics[width=6cm]{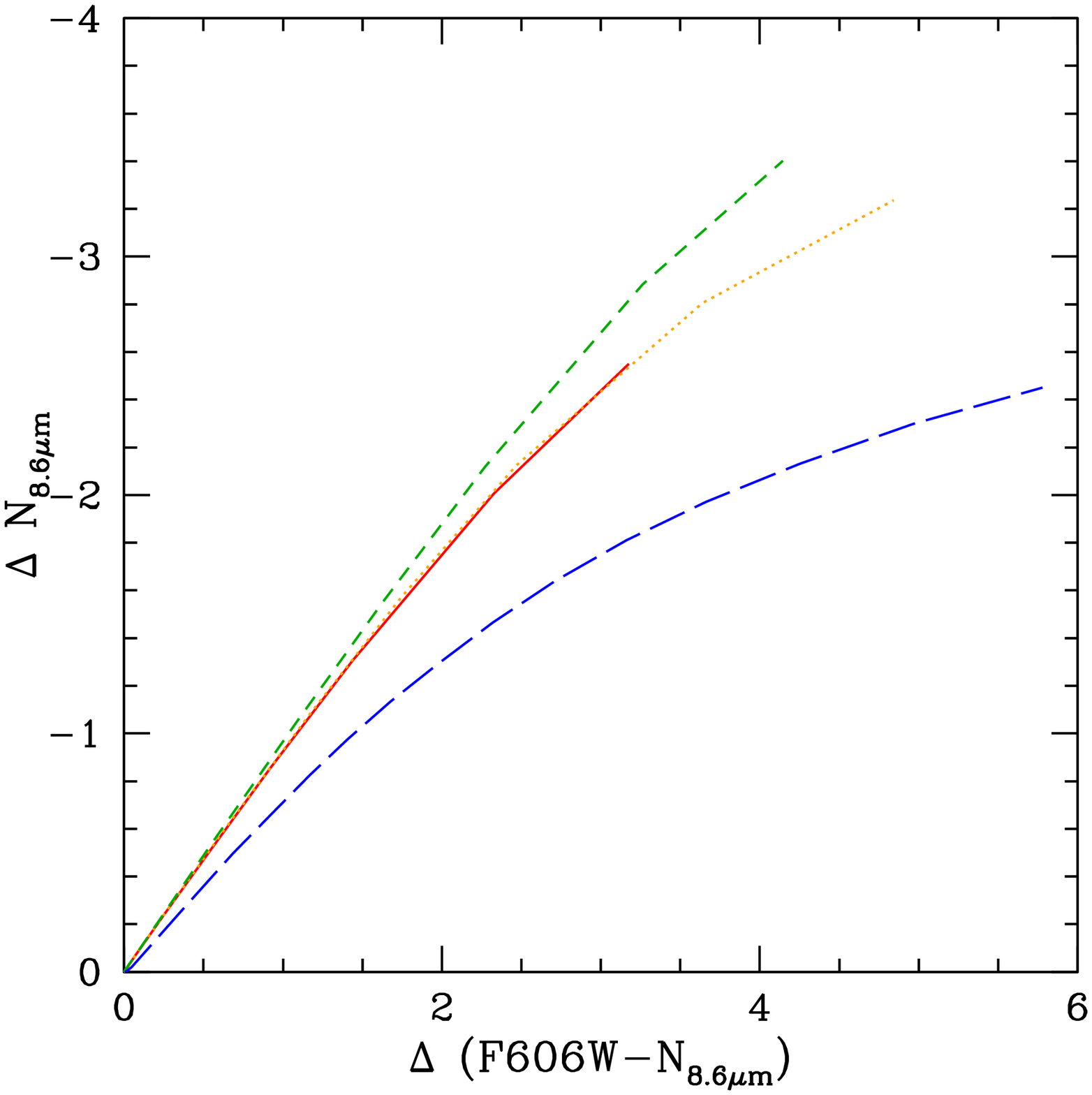}
\includegraphics[width=6cm]{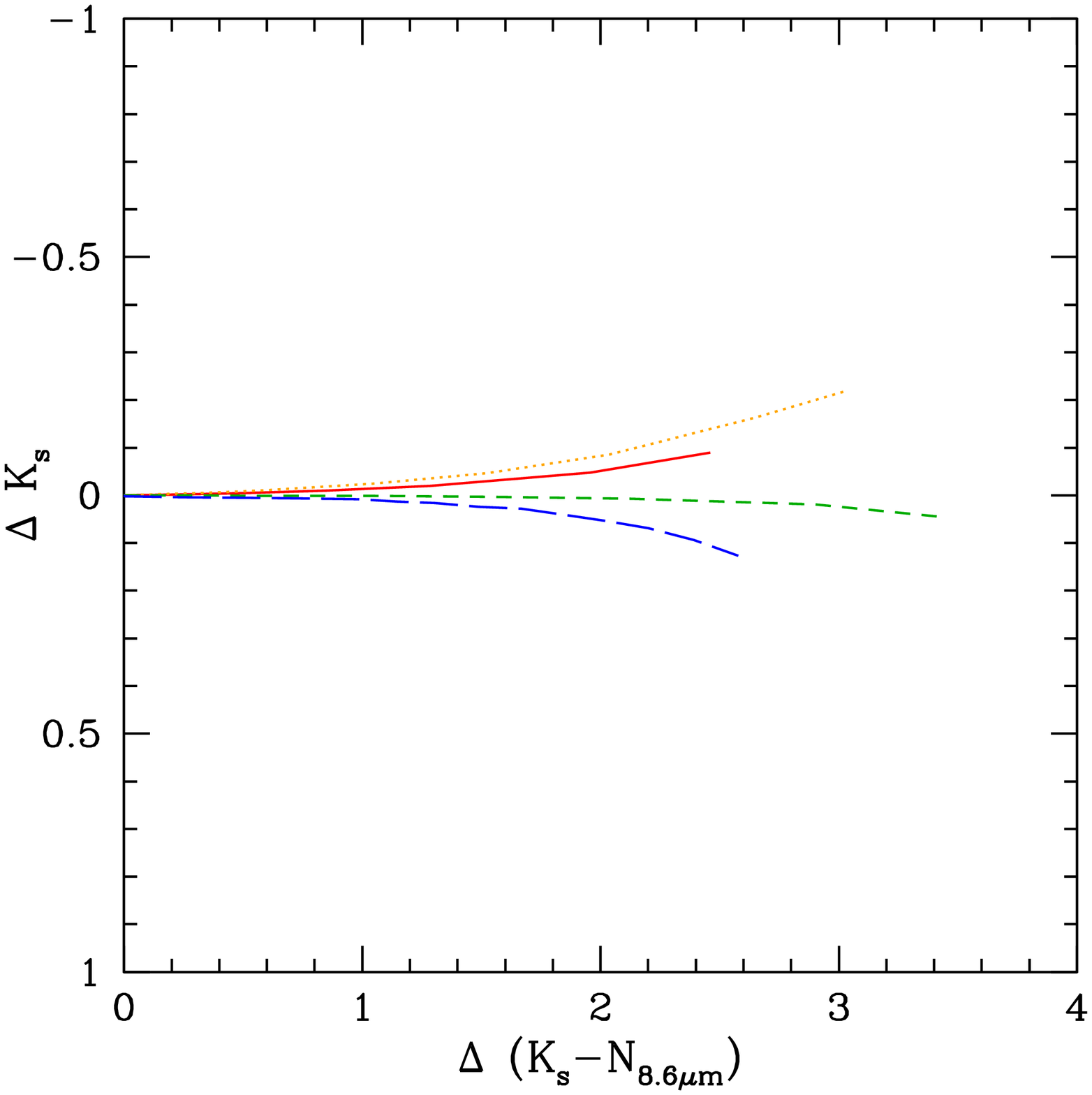}
\caption{Expected  changes in colours and magnitudes  of RGB stars
  caused by O-rich circumstellar dust  shells for an optical depth at
  $\lambda=1\micron$ varying between 0  and 1, in the CMDs $N_{\mathrm
    {8.6\micron}}$      vs.       ($m_{\mathrm      {F606W}}-N_{\mathrm
    {8.6\micron}})$      (upper     panel)     and      $K_{\rm s}$     vs.
  $(K_{\rm s}-N_{\mathrm {8.6\micron}})$ (lower panel).  The different curves are for
  different dust  compositions, namely the 100~\%  AlOx (solid), 60~\%
  silicate +  40~\% AlOx (dotted), and  100~\% silicate (short-dashed)
  mixtures from \citet{groen06},  and the silicates (long-dashed) from
  \citet{bressan98}.}
\label{f_tau1}
\end{figure}

In the previous  section we used our optical-MIR  diagram to highlight
the lack of RGB MIR-excess stars.
However,  it is  quite difficult  to  draw any  firm conclusion  based
solely on the  basis of this diagram. Indeed,  any MIR-excess emission
would  instead scatter  the  stars to  brighter  magnitudes and  redder
colours in a direction roughly parallel to the RGB morphology.
This  is quantitatively  demonstrated in  Fig.~\ref{f_tau1},  where we
plot the shift produced on the colour-magnitude diagram position of an
upper RGB star as a function of the optical depth of the circumstellar
dust shell for  several O-rich dust mixtures that  are included in the
\citet{marigo08} isochrones.
Figure ~\ref{f_tau1} shows that the  MIR excess can be best identified
using the $K_{\rm s}$,($K_{\rm s}-N_{\mathrm {8.6\micron}}$) plane (as
in the original Origlia et al.  2007 paper) instead of the $N_{\mathrm
  {8.6\micron}}$,($m_{\mathrm    {F606W}}-N_{\mathrm   {8.6\micron}}$)
plane used in the previous section.

\begin{figure}
\centering
\includegraphics[width=9cm]{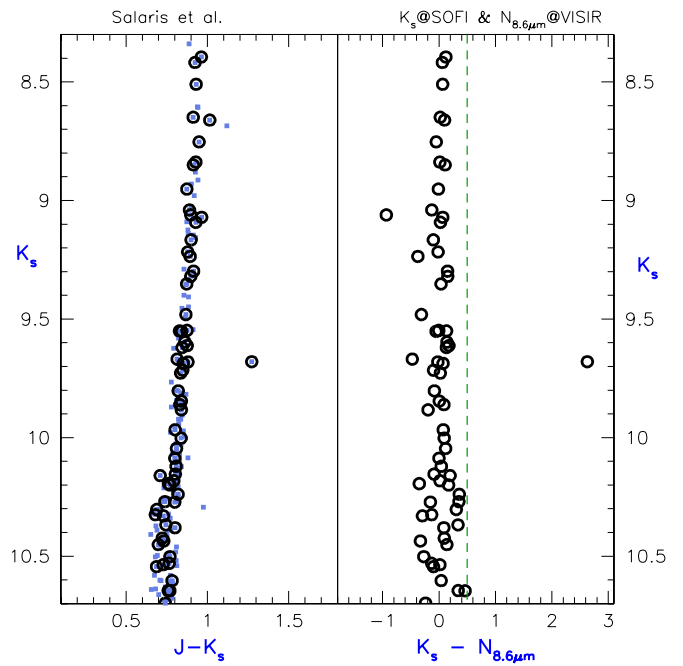}
\caption{Left-hand panel: $K_{\rm s}$,  ($J-K_{\rm s}$) SOFI@NTT based
  diagram by Salaris et al.  Light symbols display the entire catalogue,
  while open  circles highlight stars with  VISIR counterparts.  The right-hand
  panel   displays    the   SOFI@NTT   VISIR@UT3    matched   $K_{\rm s}$,
  ($K_{\rm s}-N_{\mathrm {8.6\micron}}$) diagram. 
  For  comparison,
  the vertical line at ($K_{\rm s}-N_{\mathrm {8.6\micron}}$)$=0.5$ marks the location
  of the majority  of MIR-excess stars seen in  Fig.~2 of Origlia et
  al. (2007). }
\label{f_K}
\end{figure}

To  this  end,  we  combined  the optical/MIR  catalogue  of  47Tuc  with
available NIR data and present the NIR/MIR diagrams in Fig.~\ref{f_K}.
The  left  panel displays  the  original $K_{\rm s}$,($J-K_{\rm s}$)  SOFI@NTT
based  diagram by  \citet{salaris07}.  We  remind the
reader  that this catalogue  does not  sample the  upper RGB  of 47Tuc,
 because the excellent seeing conditions basically saturated the near RGB
tip region.
The right panel of  Fig.~\ref{f_K} displays the SOFI@NTT and VISIR@UT3
matched  $K_{\rm s}$,($K_{\rm s}-N_{\mathrm {8.6\micron}}$)  diagram. 
This diagram  shows that  the RGB loci  are perfectly vertical  in the
($K_{\rm    s}-N_{\mathrm    {8.6\micron}}$)    plane,   allowing    a
straightforward   identification   of   possible   MIR-excess   stars.
Evidently, one star shows a colour excess (see Appendix~\ref{app2}).
On  the  other  hand,  Figure~2  of  Origlia et  al.   (2007)  shows  a
significant population  with MIR-excess with  ($K_{\rm s}-N_{\mathrm {8.6\micron}}$)
colours between $0.3-0.7$.
In conclusion,  our NIR-MIR diagram  shows that red giants  at $\sim2$
magnitudes from the  RGB tip and in an  interval extending for $\sim2$
more magnitudes  {\em do not} show any particular signature of
mass loss occurrence among the 47Tuc red giants.
We  note  that  \citet{iana}  have  performed a  similar  analysis  (i.e.
combining   their  re-reduced   Spitzer  data-sets   with   {\em  the}
\citealt{origlia07} dusty  sample) and found  that almost half  of the
Origlia et al.  MIR-excesses stars  {\em did not} have counterparts in
their re-reduction.
In this  regard, our  higher-resolution ground-based  NIR/MIR diagrams
leave basically no space for  the presence of a significant MIR-excess
RGB population. This ultimately  points towards the actual photometric
reduction  and intrinsic  spatial  resolution of  the  data as  the
primary cause of the detection of MIR-excess in 47Tuc.

\begin{figure*}
\centering
\includegraphics[width=14cm]{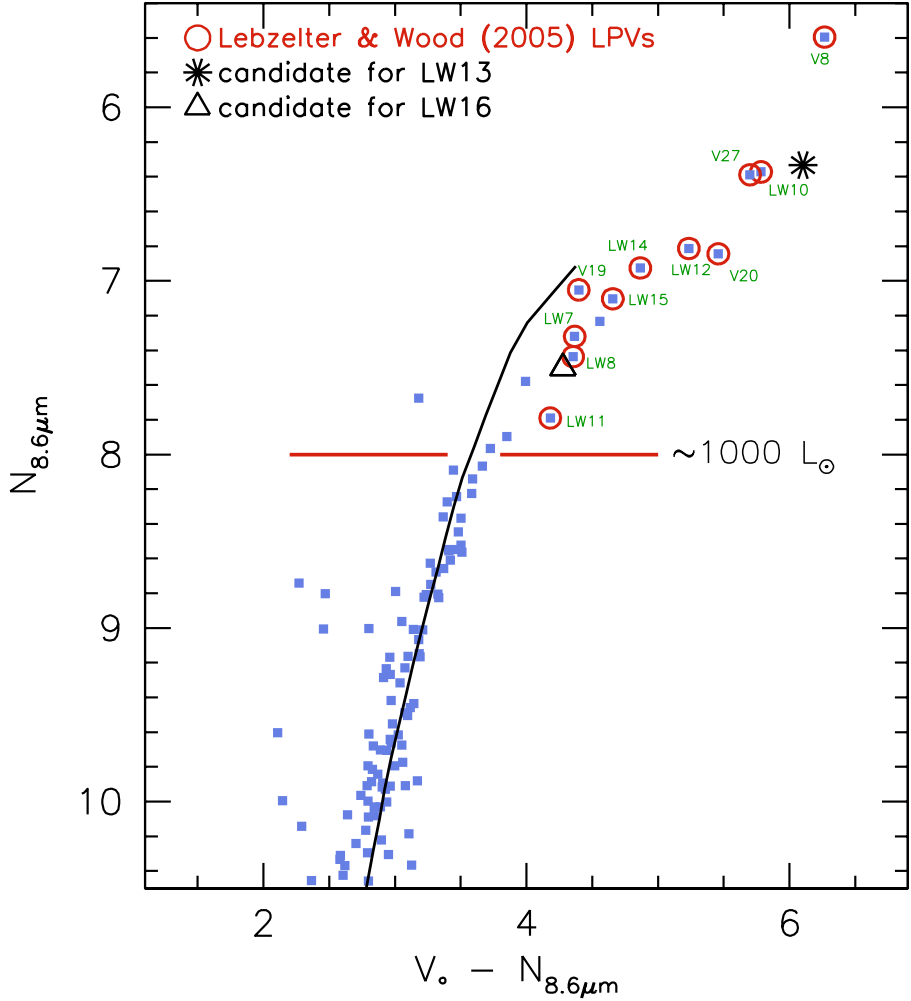}
\caption{Identified long-period variables in the $N_{\mathrm
      {8.6\micron}}$,($V_{\circ}-N_{\mathrm   {8.6\micron}}$)   plane.
    Certain  identifications  of the  LPVs  are  plotted  as red  open
    circles,  while  {\em uncertain  identifications}  are plotted  in
    different symbols, and only for the most plausible candidates.
    The $V$ magnitudes  of the LPVs were  corrected for the known
    variability cycle, i.e. {\em shifted to their average luminosity}.
    The RGB  tip is  at $N_{\mathrm {8.6\micron}}\simeq7.0$,  while the
    suggested    luminosity   {\it    break}    is   at    $N_{\mathrm
      {8.6\micron}}\simeq8.0$,  corresponding  to $\sim1000L_{\odot}$.}
\label{f_var}
\end{figure*}

\begin{figure*}
\centering
\includegraphics[width=19.5cm]{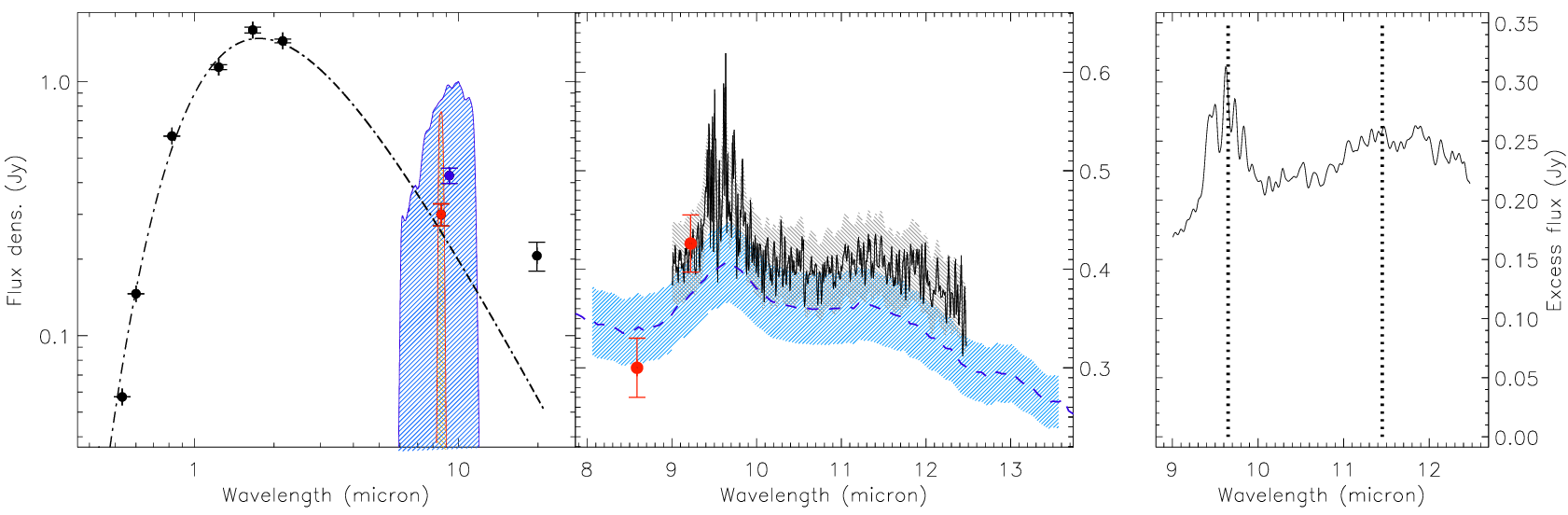}
\caption{Left-hand panel:  SED of  V8:  black filled
  circles  correspond  to   the  observed  photometry  at  $m_{\mathrm
    {F435W}}$, $m_{\mathrm {F606W}}$,  $m_{\mathrm {F814W}}$ (from the
  HST) $J$, $H$, $K_{\rm  s}$ (from 2MASS), $N_{\mathrm {8.6\micron}}$
  (from VISIR),  plus $9$ and  $18\micron$ from AKARI.   The dotted dashed
  black line is  the best-fitting black-body with  a temperature of of
  $2900K$.   The VISIR $N_{\mathrm {8.6\micron}}$  filter transmission  curve is
  shown  in red, while  that of  AKARI $9\micron$  is shown  in shaded
  blue.  The  middle panel displays  a comparison of the  VISIR (solid
  black line)  and IRS  (dashed blue line)  spectra. The  shaded areas
  highlight the  flux calibration  uncertainties estimated  for the
  two  spectra.  The  right-hand panel  shows  the MIR  excess for  V8
  (continuum  subtracted spectrum)  where  the features  at $9.7$  and
  $11.5\micron$ are highlighted with vertical dotted lines.}
\label{f_amelia}
\end{figure*}

\section{Long-period variables}
\label{s_LPV}

 Stellar  pulsation is  vital in the  production of dust  and mass
  loss (\citealt{ramdani01}).
Indeed, the luminosity  range of mass-losing red  giants is connected
to  the variable  nature  of these  stars,  and as  already argued  in
\citet[and references  therein]{iana}, dust-producing stars  tend to be
exclusively variable, while variable stars tend to produce dust.
We have identified  the list of long-period  variables presented
and examined in the \citet[2006]{leb05}  and \citet{leb06b}. 
As displayed  in Fig.~\ref{f_var}, the  majority of the LPVs  found in
our  VISIR   mosaic  has  certain  identification  (stars   in  open
circles).  In  the particular  case  of LW13,  we  note  that the  low
accuracy of  the LPV  coordinates in contrast  to the  high resolution
provided by VISIR do not  allow a definitive identification because it
falls between  two candidates stars.   There is a very  bright candidate
($N_{\mathrm {8.6\micron}}=6.333$)  at  $\sim1\farcs0$   located  west,   and  a
$N_{\mathrm {8.6\micron}}=7.233$ located at $1\farcs4$ east.
For LW16  we  have the  opposite  situation: the  LPV is  only
$0\farcs7$ away  from a $N_{\mathrm {8.6\micron}}=8.224$  candidate and $2\farcs8$
from  a brighter  $N_{\mathrm {8.6\micron}}=7.501$  candidate.
Figure~\ref{f_var},  clearly shows  that  all stars  above the  RGB tip
($N_{\mathrm {8.6\micron}}\simeq7.0$)  are AGB  LPV.  Moreover,  and  keeping in
mind the particular case of LW16, one can infer that the occurrence of
LPVs  (in the  current VISIR  coverage)  are present  down to  $\sim1$
$N_{\mathrm {8.6\micron}}$ magnitude below the RGB tip.

Following the  referee's suggestion, we also examined  the mean optical
$V$-magnitudes of the brightest and most variable AGB stars.  Indeed, these
can  show quite  significant  $V$-variations (e.g.  V8  varies up  to
$\sim1.7$  mag.)   depending  on  the epoch  of  observation,  which
affects the star's proper location in the colour-magnitude diagram.
On  the  other  hand,  one  can  assume  that  the  amplitude  of  the
$N_{\mathrm  {8.6\micron}}$  variations would  be  failry  small.   For
example, V8,  the brightest of our  AGB variables, is  believed to have
too small  $K_{\rm s}$-variations (\citealt{leb06}) for  the phase of
observation to be considered significant.

  The majority of monitoring studies that address these bright AGB stars
  in 47Tuc are ground-based (e.g.  \citealt{leb06}; \citealt{wel04}).
  We  therefore  converted   the  HST/ACS  $m_{\mathrm {F606W}}$  magnitudes  into
  Johnson-$V$,  following the procedure  outlined in  \citet{ivo08}.
  It employs the stars dereddened $m_{\mathrm {F606W}}$ magnitudes and
  $m_{\mathrm   {F606W}}-m_{\mathrm    {F814W}}$   colours   and   the
  transformation     coefficients     listed     in    Table~14     of
  \citet{sirianni05}.   The  resulting  $V_{\circ}$   and  $I_{\circ}$
  magnitudes  were  compared   to  the  47Tuc  ground-based  catalogue
  presented in \citet{alves05}, which excellently agreed.
  The mean $V$-luminosities  were then derived using the \citet{leb05}
  and \citet{leb06b}  light curves, which we compared  with the HST's
  (JD=2453807.6 epoch) newly converted $V$-magnitudes.
%

%
%
  With the exception of V20  (at $V_{\circ}=12.623$ is observed at its
  minimum  phase)  the remaining  LPVs  tend  (in different  measures)
  towards   having   redder   ($V_{\circ}-N_{\mathrm   {8.6\micron}}$)
  colours, which additinally support their MIR-excess.
  In  Section~\ref{s_isoc} we  cautioned against  emphasizing  the small
  colour   deviation  between   the  stars   and  the   isochrone,  in
  correspondence of the  $\sim1000L_{\odot}$ luminosity level. We note,
  however, that  correcting for the  optical $V$ mean  magnitudes would
    increase this off-isochrone  colour deviation (starting at
  $N_{\mathrm {8.6\micron}}\sim8.0$) even more.
 Overall, Figure~\ref{f_var} shows  that dusty mass loss: (i) is
    triggered  at the  $\sim1000L_{\odot}$ luminosity  level;  (ii) is
    strongly associated with  the RGB/AGB pulsation properties; (iii)
    is  relatively  weak along  the  RGB; and  lastly  (iv)  is not  an
    episodic process.

In  Fig.~\ref{f_amelia} we  focus our  attention on  the  brightest of
these   LPVs  in  our   sample  (V8)   and  complement   the  derived
$N_{\mathrm {8.6\micron}}$  flux  with  a VISIR  $8-13$$\micron$  low-resolution
($R\sim350$) spectrum (see \citealt{vanloon06} for similar studies).
According to the study of \citet{leb06}, V8 is a $\sim100$-day period
 fundamental-mode radial pulsator.
The  middle  panel  of  Fig.~\ref{f_amelia} displays  the  VISIR  V8
spectrum  (black  solid  line)   compared to  the  lower-resolution
(R$\sim100$)  SPITZER/IRS  spectrum  (light  blue  line)  presented  in
\citet{leb06b}.
The  shaded  areas highlight  the  flux  calibration uncertainties 
estimated for both spectra. In particular, for the IRS spectrum we 
made use of  the pipeline reduction, which  has an uncertainty of
$\sim10\%$,  in  comparison  to  $\sim5\%$ uncertainty  for  the  more
detailed reduction presented in \citet{leb06b}.
Interestingly,   the  main  dust  features in  the  VISIR and  IRS
  spectra are very  similar. There is a visible  flux level difference
  between  the VISIR and  IRS spectra,  but we  note that  this is
  within the estimated errors, as shown by the shaded areas.

For a  better characterization of its  MIR excess, we  compiled a
fairly  complete SED  of V8  using photometric  data from  HST, 2MASS,
VISIR,  and AKARI\footnote{AKARI-FIS  Bright Source  Catalogue Release
  note Version  1.0 Yamamura, I.,  Makiuti, S., Ikeda, N.,  Fukuda, Y,
  Oyabu, S,  Koga, T.,  White, G. J.,  2010}, covering the  range from
near-UV to the MIR.
The  best-fit temperature  (dotted-dashed line  in the  left  panel of
Fig.~\ref{f_amelia}) is found to  be $2900K$, and this temperature was
subsequently used to model the continuum.
At   longer   wavelength,  we   note   that   the  VISIR   $N_{\mathrm
  {8.6\micron}}$ data point  shows a better match with  respect to the
AKARI $9\micron$ data point. The  difference is probably caused by the
slightly ``bluer'' and narrower VISIR PAH1 filter, which make it less
sensitive to dust emission, while  the AKARI data are more sensitive to
the silicate feature.

To  better   characterize  the  silicate  feature,   we  employed  the
previously fitted  black-body as continuum and subtracted  it from the
dusty VISIR spectrum of V8.
We  note  that  because  we   are  in  the  Rayleigh-Jeans  domain,  the
difference  between subtracting a  black-body of  this temperature  and one
within $\pm1000K$ should be negligible.
The right  panel finally displays the VISIR  V8 continuum-subtracted
spectrum. 
 The  dominant dust features  are expected at around  $9.7$, $11.5$
  and $13\micron$.  The VISIR  V8 spectrum is limited to $12.5\micron$;
  however, the $13\micron$ feature was detected in the studies of
  \citet{leb06b} and \citet{ianb}.
  As  discussed  in  \citet{leb06b},   the  dust  ($9.7$,  $11.5$  and
  $13\micron$)  features show different  strength in  correlation with
  the star's position  along the AGB. At the  highest luminosities, as
  is the case for V8, the $9.7\micron$ feature is expected to dominate
  those at  the remaining two  wavelengths. This is clearly  visible in
  the right panel of Fig.~\ref{f_amelia}.
  In particular, the $\sim10\micron$ feature of V8 consists of a broad
  feature with  an emerging $9.7\micron$ silicate bump. 
  The feature at $11.5\micron$ is, as expected,  weaker than
  that at  $9.7\micron$.  Although there is no  general consensus (see
  \citealt{leb06b}), it is believed  that the feature at $11.5\micron$
  is caused by amorphous aluminum oxide $Al_2O_3$ (\citealt{ianb}).

  The strong  $9.7\micron$ silicate  bump observed  in the
  VISIR spectrum  was not detected  in the only other  ground-based MIR
  spectrum of  V8 presented by  \citet{vanloon06}. The absence  of this
  feature has  led \citet{leb06b} to  suggest possible phase-dependent
  background impacting the \citet{vanloon06} TIMMI2 spectrum. 
  However,  the detection  of the  feature in  our  ground-based VISIR
  spectrum clearly  point towards a higher-sensitivity  provided by the
  VISIR/VLT (8-m class telescope)  combination with respect to that of
  TIMMI2 and ESO/3.6-m telescope.

  The V8 $9.7\micron$ silicate feature places V8 as a broad+sil AGB in
  the  dust classification  scheme of  \citet{speck00}.  
  Interestingly, however, and as  evident from the spectrum presented in
  \citet{leb06b} and  \citet{ianb}, V8 displays the dust  features at
  all three wavelengths. This adds some confusion to its classification.
  Indeed, the $13\micron$ feature  (very similar to that of SAO~37673)
  is commonly  found in non-Mira semi-regular  variables according
  to   \citet{speck00}.    On    the   other   hand,    following
  \citet{sloan10}, the same $13\micron$  feature is detected in Miras.
  This  agrees  with the  results presented  in \citet{leb05},
  showing that the V8 velocity  curve resembles those of typical Miras
  found in the solar  neighbourhood.  Yet again, the $V$-band amplitude
  of  V8 of  $\sim1.7$  mag.  (\citealt{arp63})  is  smaller than  the
  nominal $2.5$ mag. value used in the classical definition of Miras.
  In  conclusion,  a  proper  classification of  V8  is  intrinsically
  difficult. Future monitoring  of the V8 (and other  dusty AGB stars)
  silicate  feature is  required to  better characterize  the relation
  between dust production and pulsation modes.

\section{Conclusions}
\label{s_conc}
The Spitzer-based  \citet{origlia07} study  of 47Tuc
has triggered a  wave of interest concerning the  process of mass loss
in red giants.
Their  results showed  that a  significant fraction  of the  47Tuc RGB
population  has  MIR-excess. In  particular,  this MIR-excess  was
affecting red giants already at the horizontal branch level.
This initiated a  very interesting debate on whether  the  Spitzer
  results were  induced by its low  spatial resolution and
the extreme crowding  of the cluster core.

The VISIR high-resolution capabilities have allowed us to address this
debate  with  an  independent  data set that  provided  the  cleanest
possible catalogue to test the Spitzer-based results.
We have complemented our VISIR $8.6\micron$ data with near-UV, optical
and NIR  photometry  to construct  colour-magnitude diagrams
best-suited to detect the MIR excess along the RGB.
From  the analysis  of  these  diagrams and  sampling  down to  about
$\sim3$ magnitudes below the RGB tip,  we found no evidence for the
presence of dusty circumstellar envelopes around 47Tuc RGB stars.
In particular, down  to this magnitude level, all  stars (but one) are
consistent   with  no   MIR-excess. 
Fainter than  $N_{\mathrm {8.6\micron}}\simeq10$ (i.e. $\ge4.0$  mag.  below the
RGB tip)  photometric uncertainties do not allow  any firm conclusion,
but the data are still consistent with the null-hypothesis.
In  conclusion,  we do  not  confirm the  results  of  Origlia et  al.
(2007). Instead, we show that RGB stars are not affected by dusty mass
loss   except those within  $\sim1$ magnitude range from the
RGB tip.

  There is  an observed  {\em break}  at $N_{\mathrm {8.6\micron}}\simeq8.0$
  that signals the  onset of a colour deviation  between the stars and
  the  theoretical  isochrone. 
  Uncertainties in  the colour transformations and  opacity and mixing
  length  treatment   all  caution  against   emphasizing  this  colour
  deviation between  the data and  the isochrone. Nevertheless,  it is
  interesting  to  note  that  this luminosity  level  corresponds  to
  $\sim1000L_{\odot}$,  which was  suggested  by   independent  determinations
  (\citealt{ianb}) to sign the onset of the dusty mass loss.

Finally, the  VISIR high-resolution capabilities that were  demonstrated in this
paper will allow  us to plan other studies of  the brightest giants in
Galactic globular clusters, spanning a wide range in metallicity.
This  and related projects  will surely  benefit from  the forthcoming
ESO/VISIR detector upgrade, which is planned for  October 2012.

\begin{acknowledgements}
  We thank  the anonymous referee  for his/her comments  that improved
  the  presentation of  the paper.   We  also thank  Jay Anderson  and
  Maurizio Salaris for  providing us with the HST  and NTT catalogues,
  respectively.   Caroline   Foster   and  Magaretha   Pretorius   are
  acknowledged for  a careful  reading of the  manuscript.  LG  and AB
  acknowledge partial support  from contract ASI-INAF I/009/10/0. This
  research is  based on observations  with AKARI, a JAXA  project with
  the participation of ESA.

\end{acknowledgements}

\clearpage
%

\longtab{3}{
\begin{longtable}{cccccccccccl}
  \caption{The VISIR  $N_{\mathrm {8.6\micron}}$ catalogue  of the 47Tuc
    central $1\farcm15$  region, along with  the $m_{\mathrm {F606W}}$
    and $m_{\mathrm {F814W}}$ magnitudes (from Anderson et al.  2009),
    the  $m_{\mathrm {F336W}}$  and  $m_{\mathrm {F435W}}$  magnitudes
    (reduced in this paper), the  $J$- and $K$-magnitudes (from Salaris
    et  al.    2007),  the  flux   value  in  Jansky  and   known  LPV
    identification.
\label{t_xonline}
}\\
\hline\hline
ID & RA$_{\mathrm {J2000}}$ & Dec.$_{\mathrm {J2000}}$ & $m_{\mathrm {F606W}}$ & $m_{\mathrm {F814W}}$ & $m_{\mathrm {F336W}}$ & $m_{\mathrm {F435W}}$ & $N_{\mathrm {8.6\micron}}$ & $J$ & $K$ & Jansky & LPV \\
\endfirsthead
\multicolumn{5}{c}{{\tablename} \thetable{} -- \scriptsize Continued} \\
\hline\hline
ID & RA$_{J2000}$ & Dec.$_{J2000}$ & $m_{F606W}$ & $m_{F814W}$ & $m_{F336W}$ & $m_{F435W}$ & $N_{8.6\micron}$ &  $J$ & $K$ &Jansky & LPV \\
\hline\hline
\endhead
\hline\hline
1 & 6.03575417 & $-$72.06523056 & 10.950 & 9.037 & --- & 13.779 & 5.595 &	 ---      &  ---         &	 0.286 & V8 \\ 
2 & 6.03217083 & $-$72.07544167 & 11.658 & 9.401 & --- & 14.338 & 6.333 &     ---   &      ---     &	 	0.145 & LW13? \\ 
3 & 6.01042083 & $-$72.08541111 & 11.486 & 9.347 & 16.259 & 14.196 & 6.372 &      ---  &   ---        &	 	0.140 & LW10 \\ 
4 & 6.06311667 & $-$72.07680278 & 11.178 & 8.671 & --- & 13.898 & 6.388 &      ---  &   ---        &	 	0.138 & V27 \\ 
5 & 6.01663333 & $-$72.08614167 & 11.498 & 9.595 & 16.444 & 13.688 & 6.813 &      --- &   ---        &	 	0.093 & LW12 \\ 
6 & 6.06023750 & $-$72.08592778 & 12.193 & 9.920 & 16.314 & 14.159 & 6.844 &      --- &   ---        &	 	0.090 & V20 \\ 
7 & 6.03921250 & $-$72.08029167 & 11.094 & 9.625 & 16.101 & 13.259 & 6.926 &      ---  &   ---        &	 	0.084 & LW14 \\ 
8 & 6.06149167 & $-$72.07903611 & 11.120 & 9.765 & 14.757 & 13.475 & 7.052 & 10.953  & 9.680  &		0.075 & V19 \\ 
9 & 6.04650417 & $-$72.08596389 & 11.213 & 9.663 & 16.147 & 13.555 & 7.103 &      ---  &   ---        &	 	 0.071 & LW15 \\ 
10 & 6.03425417 & $-$72.07556667 & 11.478 & 9.853 & 16.127 & 13.501 & 7.233 &  ---    &  ---       &	 	 0.063 & LW13? \\ 
11 & 5.98718750 & $-$72.09240278 & 11.253 & 9.800 & 15.808 & 13.470 & 7.320 &  ---    &  ---       &	 	 0.058 & LW7 \\ 
12 & 5.98984583 & $-$72.09153333 & 11.368 & 9.916 & 16.570 & --- & 7.437 &  ---    &  ---       &	 	 0.052 & LW8 \\ 
13 & 6.05426667 & $-$72.08134722 & 11.358 & 9.917 & 15.919 & 13.418 & 7.501 &  ---    &  ---       &	 	 0.049 & LW16? \\ 
14 & 6.03235833 & $-$72.08616944 & 11.320 & 9.964 & --- & 13.305 & 7.579 &  ---    &  ---       &	 	 0.046 & --- \\ 
15 & 6.02616250 & $-$72.07928333 & 10.622 & 9.316 & --- & 13.429 & 7.677 &  ---    &  ---       &	  	 0.042 & --- \\ 
16 & 6.01315417 & $-$72.08076389 & 11.465 & 10.086 & --- & 13.475 & 7.790 & ---     & ---        &	 	 0.038 & LW11 \\ 
17 & 6.03469167 & $-$72.08076667 & 11.503 & 10.163 & 15.911 & 13.493 & 7.898 & ---     & ---        &	      	 0.034 & --- \\ 
18 & 6.00691250 & $-$72.08238889 & 11.461 & 10.179 & --- & 13.474 & 7.965 & ---     & ---        &	 	 0.032 & --- \\ 
19 & 6.00600000 & $-$72.08996944 & 11.518 & 10.299 & 15.766 & 13.548 & 8.067 & ---     & ---        &	 	 0.029 & --- \\ 
20 & 6.04209167 & $-$72.08168611 & 11.348 & 10.228 & --- & 13.460 & 8.089 & ---     & ---        &	 	 0.029 & --- \\ 
21 & 5.99916667 & $-$72.09126944 & 11.511 & 10.277 & 15.845 & --- & 8.140 & ---     & ---        &	 	 0.027 & --- \\ 
22 & 6.05709167 & $-$72.08125278 & 11.621 & 10.497 & 15.760 & 13.552 & 8.224 & ---     & ---        &	 	 0.025 & LW16? \\ 
23 & 6.01279167 & $-$72.08196667 & 11.511 & 10.337 & 15.621 & 13.479 & 8.243 & ---     & ---        &	 	 0.025 & --- \\ 
24 & 6.00649167 & $-$72.07952222 & 11.483 & 10.348 & --- & 13.493 & 8.274 & 9.358  & 8.395 &		 0.024 & --- \\ 
25 & 6.02124167 & $-$72.08179167 & 11.540 & 10.419 & 15.643 & 13.431 & 8.359 & 9.342   & 8.418  &		 0.022 & --- \\ 
26 & 6.07743750 & $-$72.07876667 & 11.655 & 10.435 & 15.795 & 13.645 & 8.367 &  ---    &   ---      &	 	 0.022 & --- \\ 
27 & 6.01575417 & $-$72.07908333 & 11.740 & 10.608 & --- & 13.627 & 8.446 & 9.441  & 8.510  &		 0.021 & --- \\ 
28 & 6.03135000 & $-$72.09172500 & 11.844 & 10.740 & 15.598 & 13.700 & 8.522 & ---     & ---        &	 	 0.019 & --- \\ 
29 & 6.03265833 & $-$72.08903611 & 11.834 & 10.712 & 15.777 & 13.756 & 8.549 & ---     & ---        &	 	 0.019 & --- \\ 
30 & 6.01141250 & $-$72.08412778 & 11.774 & 10.664 & --- & 13.648 & 8.549 & ---     & ---        &	 	 0.019 & --- \\ 
31 & 6.02700000 & $-$72.09286111 & 11.782 & 10.660 & 15.586 & 13.637 & 8.556 & ---     & ---        &	 	 0.019 & --- \\ 
32 & 6.04496667 & $-$72.08403889 & 11.881 & 10.740 & 15.920 & 13.806 & 8.563 & 9.676 & 8.661 &		 0.019 & --- \\ 
33 & 6.03616250 & $-$72.08653889 & 11.851 & 10.745 & 15.602 & 13.690 & 8.610 & ---     & ---        &	 	 0.018 & --- \\ 
34 & 6.03457917 & $-$72.07068611 & 11.722 & 10.640 & --- & 13.514 & 8.628 & 9.562 & 8.649 &		 0.017 & --- \\ 
35 & 6.04270417 & $-$72.09281667 & 11.869 & 10.840 & 15.409 & --- & 8.656 & ---     &  ---       &	 	 0.017 & --- \\ 
36 & 5.99227083 & $-$72.09194722 & 11.810 & 10.705 & 15.721 & 13.593 & 8.677 & ---     &  ---       &	 	 0.017 & --- \\ 
37 & 6.01921250 & $-$72.07953889 & 10.914 & 10.067 & --- & 13.622 & 8.742 & 9.764 & 8.850 &		 0.016 & --- \\ 
38 & 6.08285833 & $-$72.08009444 & 11.825 & 10.664 & --- & --- & 8.750 &  ---    & ---        &	 	 0.016 & LW18? \\ 
39 & 6.00770833 & $-$72.06926111 & 11.590 & 10.406 & --- & 13.547 & 8.789 &  ---    & ---        &	 	 0.015 & --- \\ 
40 & 6.03491250 & $-$72.07908333 & 11.127 & 10.136 & --- & 13.719 & 8.803 & 9.702 & 8.753 &		 0.015 & --- \\ 
41 & 6.03509167 & $-$72.07666111 & 11.962 & 10.896 & 15.637 & 13.769 & 8.805 &  ---    &  ---       &	 	 0.015 & --- \\ 
42 & 5.99954167 & $-$72.08896667 & 11.860 & 10.736 & 15.738 & 13.714 & 8.807 &  ---    &  ---       &	 	 0.015 & --- \\ 
43 & 6.07183750 & $-$72.07508333 & 11.874 & 10.800 & 15.581 & 13.693 & 8.823 & 9.767 & 8.838 &		 0.015 & --- \\ 
44 & 5.98394167 & $-$72.07970000 & 11.993 & 10.938 & --- & 13.753 & 8.825 &  ---    & ---        &	 	 0.015 & --- \\ 
45 & 6.01211667 & $-$72.07690000 & 11.859 & 10.836 & --- & 13.617 & 8.963 & 9.826 & 8.952 &		 0.013 & --- \\ 
46 & 6.01251250 & $-$72.08536389 & 11.666 & 10.700 & --- & 13.257 & 9.002 &  ---    &  ---       &	 	 0.012 & --- \\ 
47 & 6.03417500 & $-$72.07883056 & 11.297 & 10.251 & --- & 13.877 & 9.005 & 10.036 & 9.071 &		 0.012 & --- \\ 
48 & 6.02413333 & $-$72.08595556 & 11.977 & 10.908 & 15.579 & 13.767 & 9.008 & ---	  &  ---       &	  	 0.012 & --- \\ 
49 & 6.02750417 & $-$72.07719722 & 12.068 & 11.050 & --- & 13.820 & 9.010 & ---  & ---        &	 	 0.012 & --- \\ 
50 & 6.06523333 & $-$72.07621389 & 12.083 & 11.036 & 15.654 & 13.864 & 9.066 & 10.021 & 9.092 &		 0.012 & --- \\ 
51 & 6.00500833 & $-$72.08565000 & 12.182 & 11.179 & 15.544 & 13.890 & 9.147 & 10.213 & 9.298 &		 0.011 & --- \\ 
52 & 6.02227083 & $-$72.07823889 & 12.108 & 11.088 & 15.595 & 13.829 & 9.163 &  ---    &  ---       &	 	 0.011 & --- \\ 
53 & 6.04552500 & $-$72.08190833 & 12.206 & 11.205 & 15.640 & 13.953 & 9.165 & 10.218 & 9.320 &		 0.011 & --- \\ 
54 & 6.03258333 & $-$72.08553056 & 11.983 & 10.994 & 15.285 & 13.949 & 9.168 & 9.930 & 9.040 &		 0.011 & --- \\ 
55 & 5.98827500 & $-$72.08909444 & 12.177 & 11.239 & 15.367 & 13.776 & 9.230 & ---     & ---        &	 	 0.010 & --- \\ 
56 & 6.06100417 & $-$72.07429444 & 12.025 & 11.040 & 15.272 & 13.706 & 9.234 & 10.096 & 9.217&		 0.010 & --- \\ 
57 & 5.99231667 & $-$72.07389444 & 12.081 & 11.080 & --- & 13.832 & 9.267 & 10.066 & 9.166&		 0.010 & --- \\ 
58 & 5.99650417 & $-$72.08514722 & 12.060 & 11.091 & --- & 13.722 & 9.285 &  ---    &  ---       &	 	 0.010 & --- \\ 
59 & 6.04674167 & $-$72.08108333 & 12.210 & 11.225 & 15.558 & 13.898 & 9.316 & 10.225 & 9.352&		 0.009 & --- \\ 
60 & 5.99253750 & $-$72.07809444 & 12.266 & 11.349 & --- & 13.785 & 9.417 & 10.379 & 9.550&		 0.008 & --- \\ 
61 & 6.02422500 & $-$72.07881667 & 12.435 & 11.449 & 15.632 & 14.071 & 9.436 & 10.488 & 9.612&		 0.008 & --- \\ 
62 & 5.98800000 & $-$72.07528611 & 12.433 & 11.462 & --- & 14.070 & 9.457 & 10.459 & 9.597&		 0.008 & --- \\ 
63 & 5.97227500 & $-$72.07725278 & 12.427 & 11.463 & --- & 14.085 & 9.489 & 10.465 & 9.620&		 0.008 & --- \\ 
64 & 6.01190000 & $-$72.08556944 & 12.460 & 11.487 & 15.622 & 14.091 & 9.505 &  ---    & ---        &	 	 0.008 & --- \\ 
65 & 5.97001250 & $-$72.07591944 & 12.391 & 11.408 & --- & 14.038 & 9.553 & 10.423 & 9.548 &		 0.007 & --- \\ 
66 & 6.04413750 & $-$72.07805278 & 11.583 & 10.640 & --- & 13.863 & 9.604 & 10.394 & 9.552&		 0.007 & --- \\ 
67 & 6.02763333 & $-$72.06923056 & 12.262 & 11.256 & --- & 13.955 & 9.611 & 10.130 & 9.236&		 0.007 & --- \\ 
68 & 5.98771667 & $-$72.07378611 & 12.492 & 11.489 & --- & 14.125 & 9.616 & 10.539 & 9.687&		 0.007 & --- \\ 
69 & 6.02910833 & $-$72.09338611 & 12.465 & 11.493 & 15.627 & 14.101 & 9.642 & ---     & ---        &	  0.007 & --- \\ 
70 & 6.04260833 & $-$72.07754444 & 12.494 & 11.528 & 15.682 & 14.142 & 9.662 & ---     & ---        &	 	 0.007 & --- \\ 
71 & 6.04708333 & $-$72.09325556 & 12.599 & 11.664 & 15.656 & 14.199 & 9.674 & ---     & ---        &	 	 0.007 & --- \\ 
72 & 6.01968333 & $-$72.08701667 & 12.375 & 11.405 & 15.560 & 13.995 & 9.679 & ---     & ---        &	 	 0.007 & --- \\ 
73 & 6.05734583 & $-$72.07459722 & 12.461 & 11.513 & 15.578 & 14.054 & 9.702 & 10.562 & 9.681&		 0.006 & --- \\ 
74 & 5.97875417 & $-$72.07548611 & 12.513 & 11.569 & --- & 14.119 & 9.705 & 10.564 & 9.728&		 0.006 & --- \\ 
75 & 6.01417083 & $-$72.07967500 & 12.691 & 11.712 & --- & 14.175 & 9.774 & 10.692 & 9.861&		 0.006 & --- \\ 
76 & 6.01170417 & $-$72.08277500 & 12.445 & 11.462 & 15.622 & 14.108 & 9.793 & 10.349 & 9.481&		 0.006 & --- \\ 
77 & 6.02921250 & $-$72.08914722 & 12.664 & 11.727 & 15.573 & 14.239 & 9.795 &  ---    &    ---     &	 	 0.006 & --- \\ 
78 & 6.00641250 & $-$72.08320556 & 12.504 & 11.536 & 15.613 & 14.109 & 9.813 & 10.565 & 9.716&		 0.006 & --- \\ 
79 & 6.02797500 & $-$72.08624722 & 12.577 & 11.623 & 15.631 & 14.169 & 9.841 & 10.686 & 9.846&		 0.006 & --- \\ 
80 & 6.02368750 & $-$72.07871389 & 12.919 & 11.976 & 15.697 & 14.462 & 9.880 & 11.059& 10.239&		 0.006 & --- \\ 
81 & 5.99536667 & $-$72.07863611 & 12.571 & 11.605 & --- & 14.178 & 9.885 & 10.624 & 9.803&		 0.005 & --- \\ 
82 & 6.01773750 & $-$72.07815278 & 12.676 & 11.761 & --- & 14.206 & 9.892 & ---     & ---        &	 	 0.005 & --- \\ 
83 & 6.02350000 & $-$72.08798056 & 12.871 & 11.972 & 15.569 & 14.374 & 9.907 & ---     & ---        &	 	 0.005 & --- \\ 
84 & 6.04767083 & $-$72.09880833 & 12.563 & 11.611 & 15.719 & 14.178 & 9.909 & ---      & ---        &		 0.005 & --- \\ 
85 & 6.02364583 & $-$72.07712778 & 12.753 & 11.837 & --- & 14.290 & 9.911 & 10.841 &10.002&		 0.005 & --- \\ 
86 & 6.00230833 & $-$72.07988889 & 12.719 & 11.864 & --- & 14.120 & 9.917 & 11.009 &10.270&		 0.005 & --- \\ 
87 & 5.98961667 & $-$72.08073611 & 12.733 & 11.814 & --- & 14.235 & 9.929 & 10.856 &10.046&		 0.005 & --- \\ 
88 & 6.01660000 & $-$72.08079444 & 12.613 & 11.787 & --- & 14.170 & 9.964 & 10.870 &10.160&		 0.005 & --- \\ 
89 & 6.01545000 & $-$72.07172222 & 11.978 & 10.938 & --- & 13.705 & 9.995 & 9.959  &9.061&		 0.005 & --- \\ 
90 & 6.02557500 & $-$72.07643611 & 12.694 & 11.846 & --- & 14.072 & 9.997 & 10.992 &10.303&		 0.005 & --- \\ 
91 & 6.03665000 & $-$72.09245556 & 12.821 & 11.912 & 15.659 & --- & 10.002 & ---	     &  ---       &	 0.005 & --- \\ 
92 & 6.02671667 & $-$72.08566667 & 12.797 & 11.876 & 15.407 & 14.293 & 10.030 & 10.971& 10.200&		 0.005 & --- \\ 
93 & 6.02381667 & $-$72.08040000 & 12.777 & 11.930 & 15.264 & 14.253 & 10.031 & 11.113& 10.368&		 0.005 & --- \\ 
94 & 6.03000833 & $-$72.08183889 & 12.590 & 11.659 & --- & 14.146 & 10.077 & 10.723&  9.883&		 0.005 & --- \\ 
95 & 6.02480833 & $-$72.07857500 & 12.801 & 11.879 & 15.686 & 14.345 & 10.081 & 10.929& 10.121&		 0.005 & --- \\ 
96 & 5.98465417 & $-$72.08013889 & 12.766 & 11.853 & --- & 14.289 & 10.088 & 10.887& 10.087&		 0.005 & --- \\ 
97 & 6.00821667 & $-$72.07455833 & 12.310 & 11.391 & --- & 13.786 & 10.143 & 10.483&  9.669&		 0.004 & --- \\ 
98 & 5.98453750 & $-$72.07821667 & 12.824 & 11.916 & --- & 14.331 & 10.166 & 10.976& 10.182&		 0.004 & --- \\ 
99 & 6.04803750 & $-$72.07295833 & 13.181 & 12.303 & 15.749 & 14.635 & 10.184 & 11.415& 10.646&		 0.004 & --- \\ 
100 & 6.02972083 & $-$72.09260000 & 13.006 & 12.119 & 15.715 & --- & 10.221 &	---     & ---        &	  0.004 & --- \\ 
101 & 6.02974167 & $-$72.07715278 & 12.814 & 11.872 & 15.621 & 14.370 & 10.242 & 10.957& 10.154&		   0.004 & --- \\ 
102 & 6.07309583 & $-$72.07739167 & 12.971 & 12.075 & 15.466 & 14.436 & 10.295 & 11.181& 10.380&		   0.004 & --- \\
\hline
\hline
\end{longtable}
} 
\clearpage


%
\appendix
\section{Peculiar object \#1}
\label{app1}
 
\begin{figure}
\centering
\includegraphics[width=9cm]{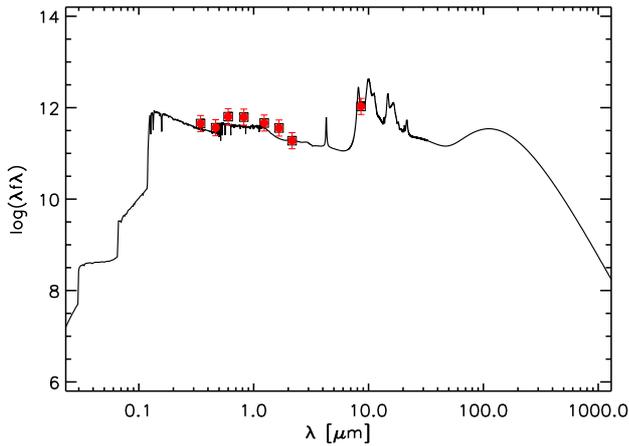}
\caption{Red  symbols  show  the  SED  of the  peculiar  VISIR  target
  (ID\#98781) with faint HST  counterpart. The black line reports the best
  fit obtained  with the MAGPHYS  code, corresponding to  a $z\sim0.3$
  redshift galaxy.}
\label{galaxy_cont}
\end{figure}

In general,  all objects reported in  Tab.\ref{t_xonline} were visible
by eye in  the VISIR images, and had an  optical HST counterpart (with
$m_{\mathrm {F606W}}\le13.0$).
  One exception  is target ID\#98781, which showed  an irregular shape
  in  the VISIR  $N_{\mathrm {8.6\micron}}$ images  and had  an  HST counterpart
  (within  a $1-$pixel  matching  radius) that  is fainter  than
  $m_{\mathrm {F606W}}\sim13.0$.
  This         target        fell        in         field\#3        at
  (RA,Dec.)$_{\mathrm {J2000}}=$($6.0323672$,$-72.0677501$) and had $m_{\mathrm {F606W}}$,
  $m_{\mathrm {F814W}}$,  $N_{\mathrm {8.6\micron}}$,  $J$-,  and  $K$-magnitudes  of
  $16.100$, $15.423$, $8.022$, $14.809$ and $14.225$, respectively.
  The irregular shape of the  target hinted at a background galaxy, and
  this  was confirmed  after  building  the SED  of  that source  (see
  Fig.~\ref{galaxy_cont}).   The SED  fitting was  performed  with the
  Multi-wavelength  Analysis of  Galaxy  Physical Properties  (MAGPHYS
  code)  presented  in \citet{cun08}.   This  indicates  a $z\sim0.3$
  redshift  galaxy  with   a  total  mass  of  $\sim2\times10^{11}
  M_{\odot}$  and  a  quite   significant  star-formation  rate  of
  $\sim150M_{\odot}$/yr.

\section{Peculiar object \#2:}
\label{app2}

Another   peculiar   red   target   (ID\#89862)  is   that   seen   in
Fig.~\ref{f_K}.          Located         at        (RA,Dec.)$_{\mathrm
  {J2000}}=$($6.06149167$,$-72.07903611$),    it    has    $m_{\mathrm
  {F606W}}$,  $m_{\mathrm {F814W}}$, $N_{\mathrm  {8.6\micron}}$, $J$-,
and  $K$-magnitudes  of  $11.120$,  $9.765$,  $7.052$,  $10.953$  and
$9.680$,   respectively.   Indeed,   it   is  V19   as   reported   in
Tab.~\ref{t_xonline}.
In the  optical $m_{\mathrm {F814W}}$,($m_{\mathrm {F606W}}-m_{\mathrm
  {F814W}}$) colour-magnitude diagram, this  star shows up as a bright
AGB candidate, located to the left  side of the RGB mean loci.  On the
other hand,  and in  the $N_{\mathrm {8.6\micron}}$  vs.  ($m_{\mathrm
  {F606W}}-N_{\mathrm {8.6\micron}}$)  diagram (see Fig.~\ref{f_var}),
it is located very close to RGB tip.
Basically, the  near-infrared $J$-  and $K$-magnitudes (as  reported in
the Salaris et al. catalogue) show a significant colour excess.
We  repeated  the  VISIR/HST/SOFI  coordinate matching  and  visually
inspected the location of this star in the images. This test confirmed
our  initial identification,  and showed  no particular  indication of
possible   blending  or   mismatch   caused by  a  faint   unresolved
companion. Moreover, the analysis on the optical $V$ mean magnitude of
this variable did not indicate significant variations.

 In  conclusion, the \citet{salaris07}  photometry of V19  shows a
  rather fainter  than expected $J$-  and $K$-magnitudes.   Indeed, the
  2MASS  $J$-   and  $K$-photometry  of  V19   ($8.695$  and  $7.567$,
  respectively) is  $\sim2$ magnitudes brighter than  that reported in
  \citet{salaris07}.  
 After excluding a possible  coordinate mismatch,
  we    caution    that    the    \citet{salaris07}    catalogue    (see
  Sec.\ref{s_satur})    suffered    photometric   saturation    around
  $K\simeq8.0$.  This throws some doubt on the \citet{salaris07} photometry
  of  V19 and, consequently,  on its  position in  the Fig.~\ref{f_K}.
  Lastly, this variable will be subject of a future investigation.

\end{document}